\documentclass[preprint,showpacs,preprintnumbers,amsmath,amssymb,aps,prd,superscriptaddress,longtable,12pt,floatfix,nofootinbib]{revtex4}

\usepackage{graphicx}
\usepackage{dcolumn}
\usepackage{bm}

\begin{document}


This version of astro-ph:0705.1315 contains the original (published) version of this article (Phys. Rev. D 76, 042008 (2007)), as well as its erratum. The original document has not been modified, but the reader should use the effective area tables from the erratum.

\pagebreak[4]

\title{Multi-year search for a diffuse flux of muon neutrinos with AMANDA-II}



\affiliation{III Physikalisches Institut, RWTH Aachen University, D-52056 Aachen, Germany}
\affiliation{Dept.~of Physics and Astronomy, University of Alaska Anchorage, 3211 Providence Dr., Anchorage, AK 99508, USA}
\affiliation{CTSPS, Clark-Atlanta University, Atlanta, GA 30314, USA}
\affiliation{Dept.~of Physics, Southern University, Baton Rouge, LA 70813, USA}
\affiliation{Dept.~of Physics, University of California, Berkeley, CA 94720, USA}
\affiliation{Institut f\"ur Physik, Humboldt-Universit\"at zu Berlin, D-12489 Berlin, Germany}
\affiliation{Lawrence Berkeley National Laboratory, Berkeley, CA 94720, USA}
\affiliation{Universit\'e Libre de Bruxelles, Science Faculty CP230, B-1050 Brussels, Belgium}
\affiliation{Vrije Universiteit Brussel, Dienst ELEM, B-1050 Brussels, Belgium}
\affiliation{Dept.~of Physics, Chiba University, Chiba 263-8522 Japan}
\affiliation{Dept.~of Physics and Astronomy, University of Canterbury, Private Bag 4800, Christchurch, New Zealand}
\affiliation{Dept.~of Physics, University of Maryland, College Park, MD 20742, USA}
\affiliation{Dept.~of Physics, Universit\"at Dortmund, D-44221 Dortmund, Germany}
\affiliation{Dept.~of Subatomic and Radiation Physics, University of Gent, B-9000 Gent, Belgium}
\affiliation{Max-Planck-Institut f\"ur Kernphysik, D-69177 Heidelberg, Germany}
\affiliation{Dept.~of Physics and Astronomy, University of California, Irvine, CA 92697, USA}
\affiliation{Dept.~of Physics and Astronomy, University of Kansas, Lawrence, KS 66045, USA}
\affiliation{Blackett Laboratory, Imperial College, London SW7 2BW, UK}
\affiliation{Dept.~of Astronomy, University of Wisconsin, Madison, WI 53706, USA}
\affiliation{Dept.~of Physics, University of Wisconsin, Madison, WI 53706, USA}
\affiliation{Institute of Physics, University of Mainz, Staudinger Weg 7, D-55099 Mainz, Germany}
\affiliation{University of Mons-Hainaut, 7000 Mons, Belgium}
\affiliation{Bartol Research Institute and Department of Physics and Astronomy, University of Delaware, Newark, DE 19716, USA}
\affiliation{Dept.~of Physics, University of Oxford, 1 Keble Road, Oxford OX1 3NP, UK}
\affiliation{Institute for Advanced Study, Princeton, NJ 08540, USA}
\affiliation{Dept.~of Physics, University of Wisconsin, River Falls, WI 54022, USA}
\affiliation{Dept.~of Physics, Stockholm University, SE-10691 Stockholm, Sweden}
\affiliation{Dept.~of Astronomy and Astrophysics, Pennsylvania State University, University Park, PA 16802, USA}
\affiliation{Dept.~of Physics, Pennsylvania State University, University Park, PA 16802, USA}
\affiliation{Division of High Energy Physics, Uppsala University, S-75121 Uppsala, Sweden}
\affiliation{Dept.~of Physics and Astronomy, Utrecht University/SRON, NL-3584 CC Utrecht, The Netherlands}
\affiliation{Dept.~of Physics, University of Wuppertal, D-42119 Wuppertal, Germany}
\affiliation{DESY, D-15735 Zeuthen, Germany}

\author{A.~Achterberg}
\affiliation{Dept.~of Physics and Astronomy, Utrecht University/SRON, NL-3584 CC Utrecht, The Netherlands}
\author{M.~Ackermann}
\affiliation{DESY, D-15735 Zeuthen, Germany}
\author{J.~Adams}
\affiliation{Dept.~of Physics and Astronomy, University of Canterbury, Private Bag 4800, Christchurch, New Zealand}
\author{J.~Ahrens}
\affiliation{Institute of Physics, University of Mainz, Staudinger Weg 7, D-55099 Mainz, Germany}
\author{K.~Andeen}
\affiliation{Dept.~of Physics, University of Wisconsin, Madison, WI 53706, USA}
\author{J.~Auffenberg}
\affiliation{Dept.~of Physics, University of Wuppertal, D-42119 Wuppertal, Germany}
\author{X.~Bai}
\affiliation{Bartol Research Institute and Department of Physics and Astronomy, University of Delaware, Newark, DE 19716, USA}
\author{B.~Baret}
\affiliation{Vrije Universiteit Brussel, Dienst ELEM, B-1050 Brussels, Belgium}
\author{S.~W.~Barwick}
\affiliation{Dept.~of Physics and Astronomy, University of California, Irvine, CA 92697, USA}
\author{R.~Bay}
\affiliation{Dept.~of Physics, University of California, Berkeley, CA 94720, USA}
% ...


\collaboration{IceCube Collaboration}
\noaffiliation


\begin{abstract}
A search for TeV -- PeV muon neutrinos from unresolved sources was performed
on \mbox{AMANDA-II} data collected between 2000 and 2003 with an equivalent
livetime of 807 days. This diffuse analysis sought to find an
extraterrestrial neutrino flux from sources with non-thermal
components. The signal is expected to have a harder spectrum than the
atmospheric muon and neutrino backgrounds. Since no excess of events was
seen in the data over the expected background, an upper limit of
\mbox{$E^{2}\Phi_\mathrm{90\% C.L.} < 7.4 \times 10^{-8}$ GeV cm$^{-2}$ s$^{-1}$
sr$^{-1}$} is placed on the diffuse flux of muon neutrinos with a
\mbox{$\Phi \propto$ E$^{-2}$} spectrum in the energy range 16 TeV to 2.5
PeV. This is currently the most sensitive \mbox{$\Phi \propto$ E$^{-2}$}
diffuse astrophysical neutrino limit. We also set upper limits for
astrophysical and prompt neutrino models, all of which have spectra
different than \mbox{$\Phi
\propto$ E$^{-2}$}.
\end{abstract}


\maketitle

\section{\label{intro}Introduction}
High energy photons have been used to paint a picture of the non-thermal
Universe, but a more complete image of the hot and dense regions of space
can potentially be obtained by combining astrophysical neutrino and gamma
ray data. Neutrinos can provide valuable information because they are
undeflected by magnetic fields and hence their paths point back to the
particle's source. Unlike photons, neutrinos are only rarely absorbed when
traveling through matter. However, their low interaction cross section also
makes their detection more challenging. The observation of astrophysical
neutrinos would confirm the predictions
\cite{wb_bound,wb_robust,wb_withredshift,sdss,sdss_revision,mpr,loeb_waxman_starburst,nellen,becker} that
neutrinos are produced in hadronic interactions in cosmic accelerators,
such as active galactic nuclei or gamma-ray bursts.

Instead of searching for neutrinos from either a specific time or location
in the sky, this analysis searches for extraterrestrial neutrinos from
unresolved sources. If the neutrino flux from an individual source is too
small to be detected by point source search techniques, it is nevertheless
possible that many sources, isotropically distributed throughout the
Universe, could combine to make a detectable signal. An excess of events
over the expected atmospheric neutrino background would be indicative of an
extraterrestrial neutrino flux.

In this paper, we report on a search for a diffuse flux of astrophysical
muon neutrinos performed with data collected by the AMANDA-II neutrino
telescope from 2000 -- 2003. To perform the search, a 5.2 sr sky region
(slightly less than 2$\pi$ sr) was monitored over a four year period, for a
total of 807 days of livetime. Before describing specifics of the analysis,
the existing diffuse neutrino models and limits and how we aim to detect
neutrinos are described in Sections \ref{section_astro_sources} and
\ref{section_detecting_nu}. In Section \ref{section_searchmethods}, typical
backgrounds to the extraterrestrial signal are discussed, as well as
how events are simulated in the detector. We also explain how an
atmospheric neutrino sample was obtained. An extensive systematic
uncertainty study is described in Section
\ref{section_systematics}. The relationship between up and downgoing events
is explored in Section \ref{section_relateupdown}. Finally, the results of
the analysis are presented in Section \ref{section_results}. Since no
excess of high energy events was seen above the predicted atmospheric
neutrino background, we set limits on the flux of extraterrestrial muon
neutrinos with a generic \mbox{$\Phi \propto $ E$^{-2}$} energy spectrum as
well as with a number of different model spectra discussed in Section
\ref{subsection_diffspectra}.

\section{\label{section_astro_sources} Astrophysical Fluxes and Limits}

The analysis presented in this paper assumes a \mbox{$\Phi \propto$
E$^{-2}$} spectrum resulting from shock acceleration processes. Although
other spectra were tested, the \mbox{$\Phi \propto$ E$^{-2}$} spectral
shape is considered a benchmark to characterize acceleration in many
sources. 

The Waxman-Bahcall upper bound \cite{wb_bound,wb_robust,wb_withredshift}
follows an \mbox{$\Phi \propto$ E$^{-2}$} spectrum and reaching below the
sensitivity of this bound has traditionally been a goal of neutrino
experiments. Nellen, Mannheim and Biermann
\cite{nellen} and Becker, Biermann and Rhode \cite{becker} have suggested
\mbox{$\Phi \propto$ E$^{-2}$} neutrino spectra with higher normalizations
than the Waxman-Bahcall bound. The other astrophysical neutrino models
tested here (Mannheim, Protheroe and Rachen \cite{mpr}, Stecker, Done,
Salamon and Sommers \cite{sdss,sdss_revision} and Loeb and Waxman
\cite{loeb_waxman_starburst}) predict different spectral shapes and are
specific predictions of neutrino fluxes from classes of objects such as
active galactic nuclei (AGN) and starburst galaxies. The models have been
derived based on a variety of astronomical results, including the observed
extragalactic cosmic ray flux and x-ray and radio measurements.

A precursor to this muon neutrino analysis was conducted with data
collected in 1997 by the \mbox{AMANDA-B10} detector \cite{diffuse97}. (In
1997, the AMANDA detector consisted of 10 sensor strings, a subset of the
19 strings in the final AMANDA-II configuration.) Other AMANDA analyses
have focused on the search for a diffuse flux of neutrinos using particle
showers or cascades \cite{cascades2000}. Cascades are caused by
$\nu_{e}$ and $\nu_{\tau}$ charged current interactions and all-flavor
neutral current interactions in the ice near the detector. Even though no
extraterrestrial signal has been detected, models can be excluded by
setting upper limits.

The Fr\'ejus \cite{frejus}, MACRO \cite{macro} and Baikal \cite{baikal})
experiments have set upper limits on the flux of astrophysical neutrinos in
the same energy region as this analysis (TeV - PeV). Published upper limits
from these experiments assuming a \mbox{$\Phi \propto$ E$^{-2}$} spectrum
are summarized along with the results of this analysis in Section
\ref{subsection_e2results}. Depending on the detector and the specific
analysis, the reported upper limit constrains either the muon neutrino flux
or the all-flavor neutrino flux. Upper limits obtained from all-flavor
analyses are not directly comparable to $\nu_{\mu}$ upper limits. However,
for a wide range of neutrino production models and oscillation parameters,
the flavor flux ratio at Earth can be approximated as 1:1:1
\cite{athar}. In that case, either a single-flavor limit can be multiplied
by three and compared to an all-flavor result, or an all-flavor limit can
be divided by three and compared to a single-flavor result.

The Baikal experiment has placed limits on models with spectra other than
\mbox{$\Phi \propto$ E$^{-2}$} \cite{baikal}, which are compared to the
results from this analysis in Section \ref{subsection_diffspectra}. Here,
nine different spectral shapes are tested, including the search for prompt
neutrinos from the decay of charmed particles. Since this analysis is
optimized on energy-dependent parameters, the optimization was performed
individually for each energy spectrum.

\section{\label{section_detecting_nu} Neutrino Detection in AMANDA}

Although chargeless particles like neutrinos are not directly observable,
the by-products of their interactions with polar ice or rock near the
detector can be observed. In particular, two types of neutrino-induced
events can be distinguished in AMANDA. All neutrino flavors can cause
hadronic or electromagnetic showers in the ice and these appear as a
momentary point-like source of Cherenkov light. Alternatively, long
track-like events are the signature of neutrino-induced muons traveling
through the detector. A cone of Cherenkov light is emitted by these muons
as they travel faster than the speed of light in ice. The present analysis
focuses exclusively on the muon track channel for identifying neutrino
events. Tau neutrinos can undergo charged current interactions and
contribute to the upgoing $\mu$ and $\nu_{\mu}$ fluxes via $\tau
\rightarrow \mu\nu_{\tau}\bar{\nu}_{\mu}$ decay. Although $\nu_{\tau}$
interactions and $\tau$ decay may contribute between 10\% to 16\% to the
E$^{-2}$ signal flux \cite{pointsource5yr}, this contribution is ignored in
this analysis.

Nineteen vertical strings hold the optical modules (OMs) for recording the
timing and position of detected photons, which is needed for reconstructing
the path of the muon \cite{nim2004}. The angular distribution between the
neutrino direction and the reconstructed muon track has a median of
2$^{\circ}$ when the highest quality events are used. The 677 OMs each
consist of a photomultiplier tube (PMT) enclosed in a pressure-resistant
glass sphere. The OMs are deployed to depths between 1500 and \mbox{2000
meters}. An event is recorded when at least 24 OMs report seeing light
within a \mbox{2.5 $\mu$s} window. AMANDA has been operating in the final
configuration with 19 strings (AMANDA-II) since 2000
\cite{nim2004}.

\section{\label{section_searchmethods}Search Methods}

This analysis uses the Earth as a filter to search for upgoing
astrophysical neutrino-induced events. The background for the analysis
consists of atmospheric muons and neutrinos created when cosmic rays
interact with Earth's atmosphere. The majority of the events registered in
the detector are atmospheric muons traveling downward through the ice.

Conventional atmospheric neutrinos arise from the decay of pions and kaons
created in cosmic ray interactions with the atmosphere. Atmospheric
neutrinos are able to travel undisturbed through the Earth. They can be
separated from atmospheric muons by their direction, namely by demanding
that the reconstructed track is upgoing. The conventional atmospheric
neutrino flux asymptotically approaches a \mbox{$\Phi \sim$ E$^{-3.7}$}
spectrum in the multi-TeV range. Prompt neutrinos are the counterpart of
the conventional atmospheric neutrino flux and will be discussed in Section
\ref{section_results}.

In the initial sample of \mbox{$5.2 \times 10^{9}$} events, many downgoing
events were misreconstructed as upgoing tracks. Misreconstruction happens
because photons scatter in the ice, causing directional and timing
information to be lost. Hence, the selected upgoing event sample not only
contains truly upgoing neutrinos, but a certain fraction of downgoing
atmospheric muons.

An energy-correlated observable was used to separate neutrino-induced
events since the predicted astrophysical neutrino flux has a much harder
energy spectrum (\mbox{$\Phi \propto$ E$^{-2}$}) than the conventional
atmospheric neutrinos from pions and kaons. Any excess of events at high
energy over the expected atmospheric neutrino background indicates the
presence of a signal.

The search method can be summarized by the following three selection steps:
\begin{enumerate}
\item Use the zenith angle from the reconstructed track to reject obviously
downgoing events.
\item Select events that have observables more consistent with typical long
upgoing tracks. This separates truly upgoing events from misreconstructed
downgoing events.
\item Use an energy-related observable (number of OMs triggered) to
separate upgoing atmospheric neutrinos from upgoing astrophysical
neutrinos.
\end{enumerate}

This analysis relied on simulated data sets of background and signal
events. Sixty-three days of downgoing atmospheric muons were simulated with
CORSIKA \cite{corsika} version 6.030 and the QGSJET01 hadronic interaction
model. The events were simulated with a \mbox{$\Phi \propto$ E$^{-2.7}$}
primary energy spectrum. These downgoing events are so frequent ($\sim$80
Hz at trigger level) that two atmospheric muon events produced by unrelated
primaries often occur in the detector during the same detector trigger
window of \mbox{2.5 $\mu$s}. Timing patterns of the light from the two
tracks may be such that the reconstruction results in a single upgoing
track. These coincident muon events may be caused by two muons which are
each individually incapable of triggering the detector with at least 24 OM
hits. However, events which only hit a few OMs can now trigger the detector
when in coincidence with another event. This means that a simple trigger
rate calculation of 80 Hz $\times$ 80 Hz $\times$ 2.5 $\mu$s is not
possible since all combinations of events with a total of at least 24 hits
can trigger the detector. These coincident muon events were simulated for
826 days of livetime and have a frequency of about $\sim$2-3 Hz at trigger
level.

Muon neutrinos with a \mbox{$\Phi \propto$ E$^{-1}$} spectrum were simulated
with \texttt{nusim} \cite{nusim} and reweighted to atmospheric neutrino
flux predictions
\cite{lipari,bartol2004,honda2004,barronlinetables,icrc2001gaisser}, as
well as to an astrophysical muon neutrino flux of \mbox{E$^{2} \Phi$ =
1$\times$10$^{-6}$ GeV cm$^{-2}$s$^{-1}$sr$^{-1}$}. The normalization of
the test signal spectrum, which is irrelevant when setting a limit, was
taken to be approximately equal the previous upper limit from
the AMANDA-B10 diffuse analysis \cite{diffuse97}.

\subsection{Filtering the Data Set}

The 2000 -- 2003 analysis covers 807 days of detector livetime between
February 2000 and November 2003.  Because of summer maintenance operations,
no data were used from the polar summer seasons. In the first stage of the
analysis, reconstruction software was used to make an initial hypothesis on
the track direction of every event based on the timing pattern of the
detected light \cite{nim2004}.

\begin{figure*}
\centering
\includegraphics*[width=3.4in]{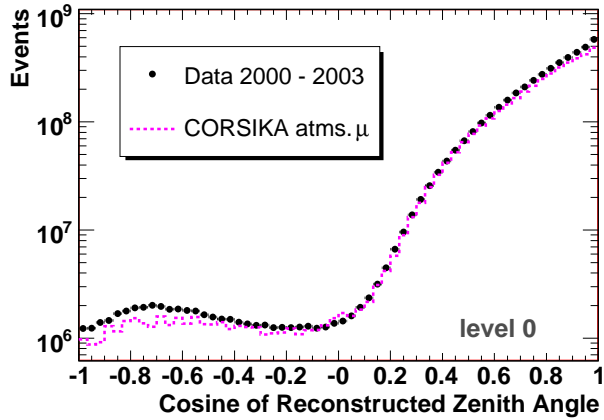}
\caption{\label{coszen_nocuts} The cosine of the reconstructed zenith angle
is shown for every event at the beginning of the analysis (level 0). The
experimental data is dominated by downgoing atmospheric muons. Events
reconstructed as upgoing appear on the left side of the plot and downgoing
events appear on the right.}
\end{figure*}

Figure \ref{coszen_nocuts} shows the zenith angle of the reconstructed
tracks for all events at the beginning of the analysis (level
0). Vertically downgoing tracks have a reconstructed zenith angle of
$0^{\circ}$ (cos($\theta$)=1). The data set was reduced to $8 \times
10^{6}$ events by removing all events reconstructed with zenith angles less
than 80$^{\circ}$ (cos($\theta$)=0.17). The remaining data set mainly
consists of misreconstructed downgoing muons and events near the horizon.

The reduction of the data by three orders of magnitude with the simple
zenith requirement made it feasible to perform more CPU-intensive track
reconstructions on the remaining events. Track parameters were adjusted to
maximize the log likelihood, given the observed light pattern. Many of the
Cherenkov photons scatter multiple times as they travel through the ice and
this changes their direction and delays the times at which they are likely
to be detected. An iterative technique was performed in which each event
was reconstructed 32 times \cite{nim2004}, each time with a different
seed. Each iteration shifts the zenith and azimuth of the track and moves
the track to pass through the center of gravity of the hits. The best track
found by the iterative search was used throughout the later stages of the
analysis.

\begin{figure*}
\centering
\includegraphics*[width=3.4in]{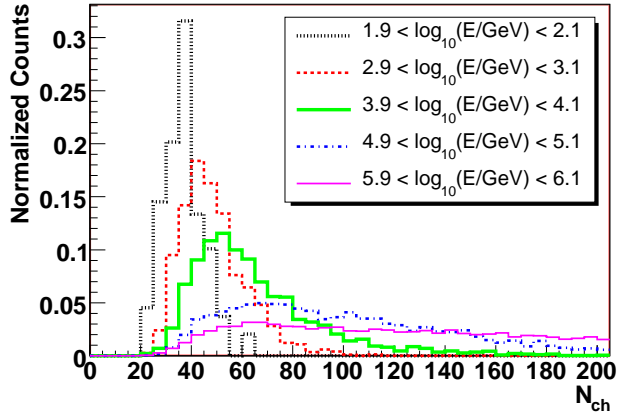}
\caption{\label{nch_vs_trueen}The number of OMs hit during an event
($N_\mathrm{ch}$) was used as an energy-correlated observable. Each line on this
$N_\mathrm{ch}$ distribution represents events with approximately the same
simulated energy. High energy events may not be contained within the
detector and hence can trigger a wide $N_\mathrm{ch}$ span.}
\end{figure*}

In order to prevent any inadvertent tuning of the event selection criteria
that would bias the result, a blindness procedure was followed which
required that further event selections were developed only on simulation
and low energy data, where the signal is negligible compared to the
background. The number of OMs triggered (from now on indicated by
$N_\mathrm{ch}$, or number of channels hit) is the energy-correlated
observable used to separate atmospheric neutrinos from astrophysical ones
(Figure \ref{nch_vs_trueen}). Only low energy data events (low
$N_\mathrm{ch}$ values) were compared to simulation. High energy data
events (high $N_\mathrm{ch}$ values) were only revealed once the final
event selection was established. Energy and $N_\mathrm{ch}$ are correlated
since high energy events release more energy in the detector causing more
hits than low energy ones. However, the correlation is not perfect since
high energy events occurring far from the detector may trigger only a few
OMs. 

Event selection was based on observables associated with the reconstructed
tracks \cite{nim2004} and is described in more detail in Appendix A. In
order to separate misreconstructed downgoing events and coincident muons
from the atmospheric and astrophysical neutrinos, events were required to
have observables consistent with long tracks and many photons with arrival
times close to those predicted for un-scattered propagation. The number of
photons arriving between $-$15 and
\mbox{+75 ns} of their expected un-scattered photon arrival time is
referred to as the number of direct hits ($N_\mathrm{dir}$). The direct
length ($L_\mathrm{dir}$) is the maximum separation of two direct hits
along the reconstructed track. The smoothness ($S$) is a measurement of how
uniformly all hits are distributed along the track and it varies between
$-$1.0 and 1.0. Positive values of the smoothness indicate more hits at the
beginning of a track and negative values indicate more hits occur toward
the end. Evenly distributed hits will have smoothness values near 0. The
median resolution ($MR$) is calculated from a paraboloid fit to the
likelihood minimum for the track
\cite{till_medres}. This method analyzes the angular resolution on an
event-by-event basis. Lastly, high quality events have higher values of the
logarithm of the up-to-down likelihood ratio, \mbox{$\Delta\mathnormal{L} =
(-\mathrm{log} \mathcal{L}_\mathrm{down}) - (-\mathrm{log}
\mathcal{L}_\mathrm{up})$}. The likelihoods $\mathcal{L}_\mathrm{up}$ and
$\mathcal{L}_\mathrm{down}$ are the product of the values of the
probability density function for the observed photon arrival times, for the
best upgoing and zenith-weighted downgoing track reconstruction
\cite{nim2004}, respectively. A more strict requirement for the likelihood
ratio was applied to vertical events than for events near the
horizon. Horizontal events tend to have smaller likelihood ratios since the
zenith angle difference between the best upgoing and zenith-weighted
downgoing track hypothesis is often small.

\begin{figure*}[!]
\centering
\includegraphics*[width=0.49\textwidth]{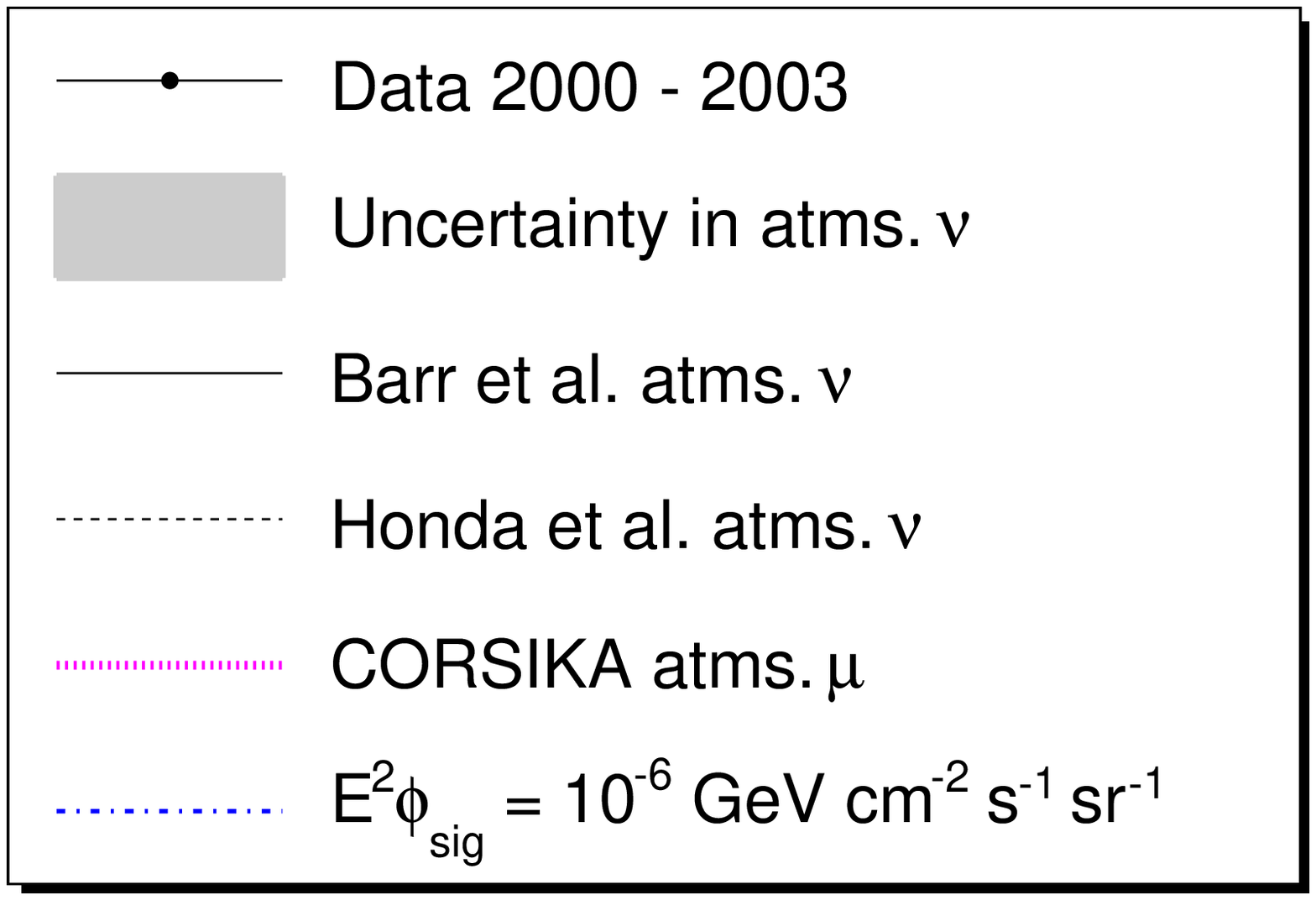}
\includegraphics*[width=0.49\textwidth]{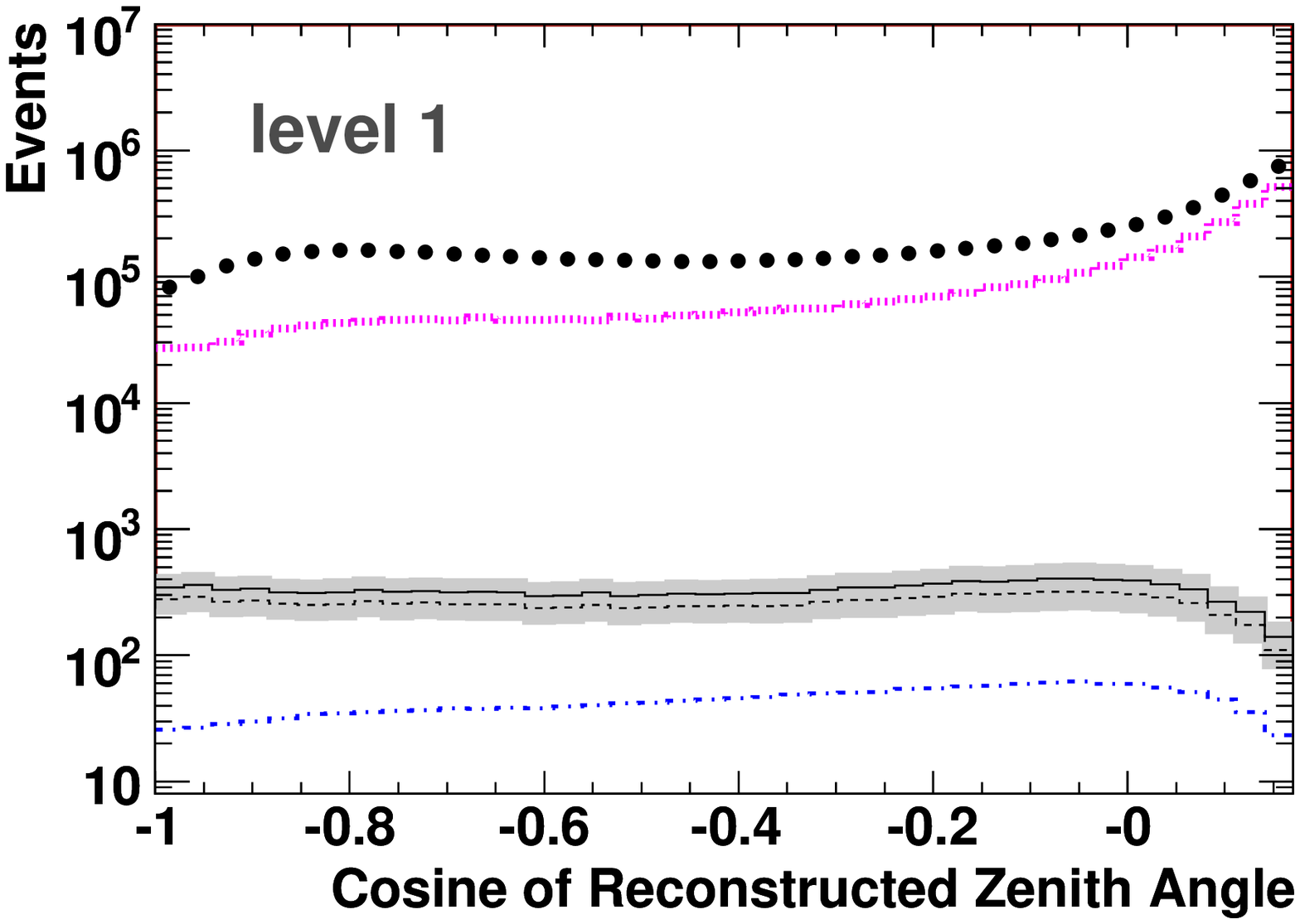}
\includegraphics*[width=0.49\textwidth]{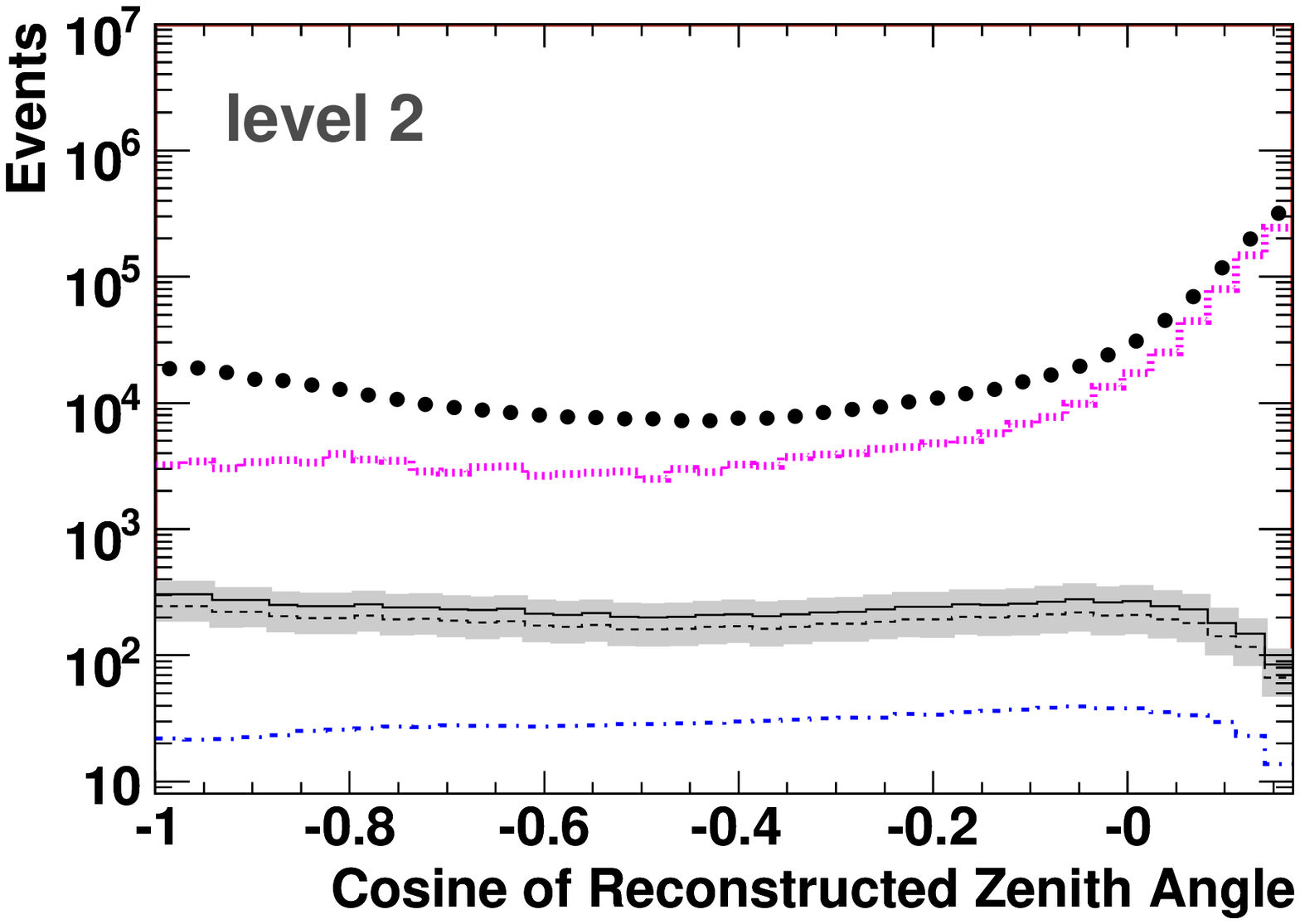}
\includegraphics*[width=0.49\textwidth]{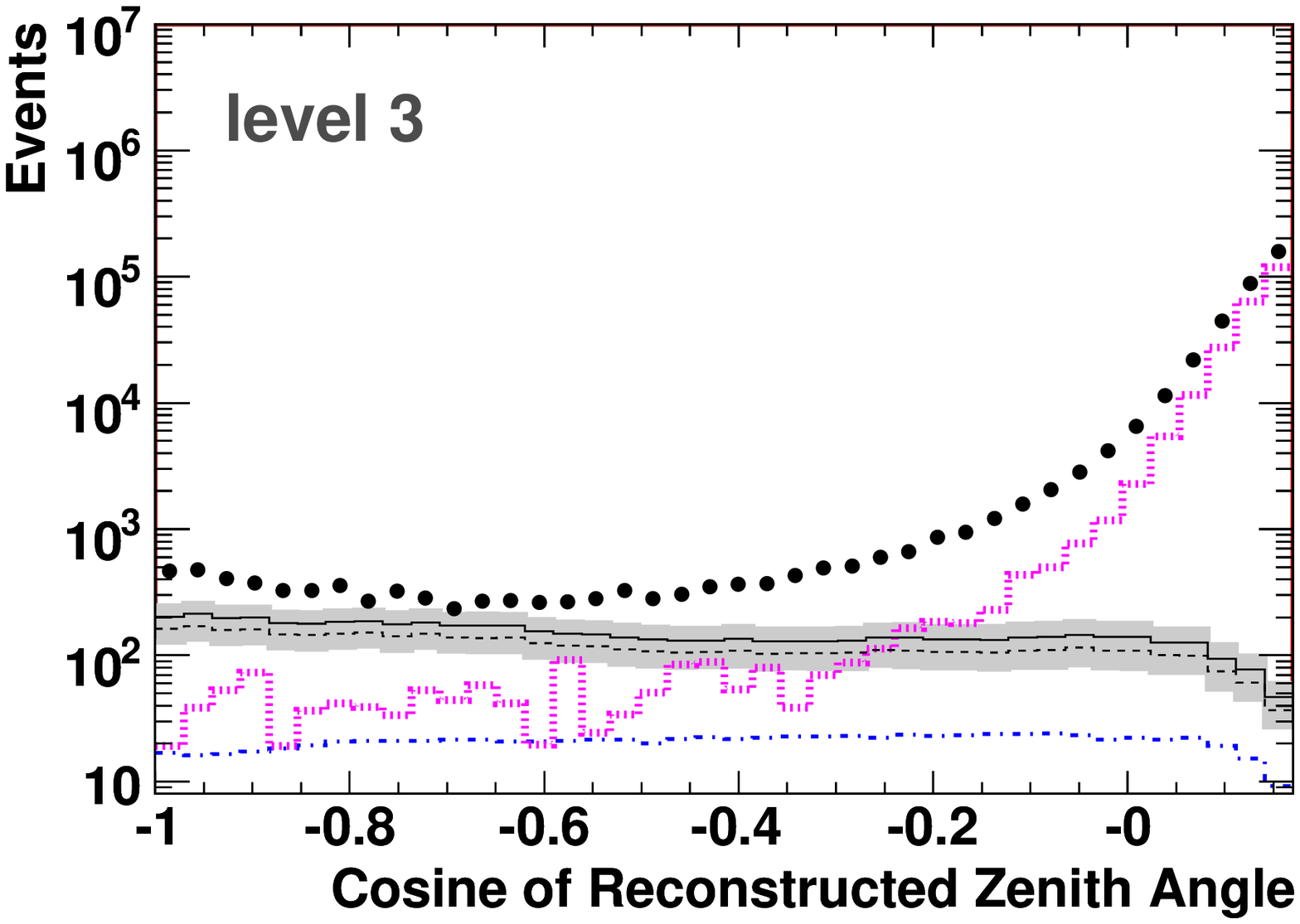}
\includegraphics*[width=0.49\textwidth]{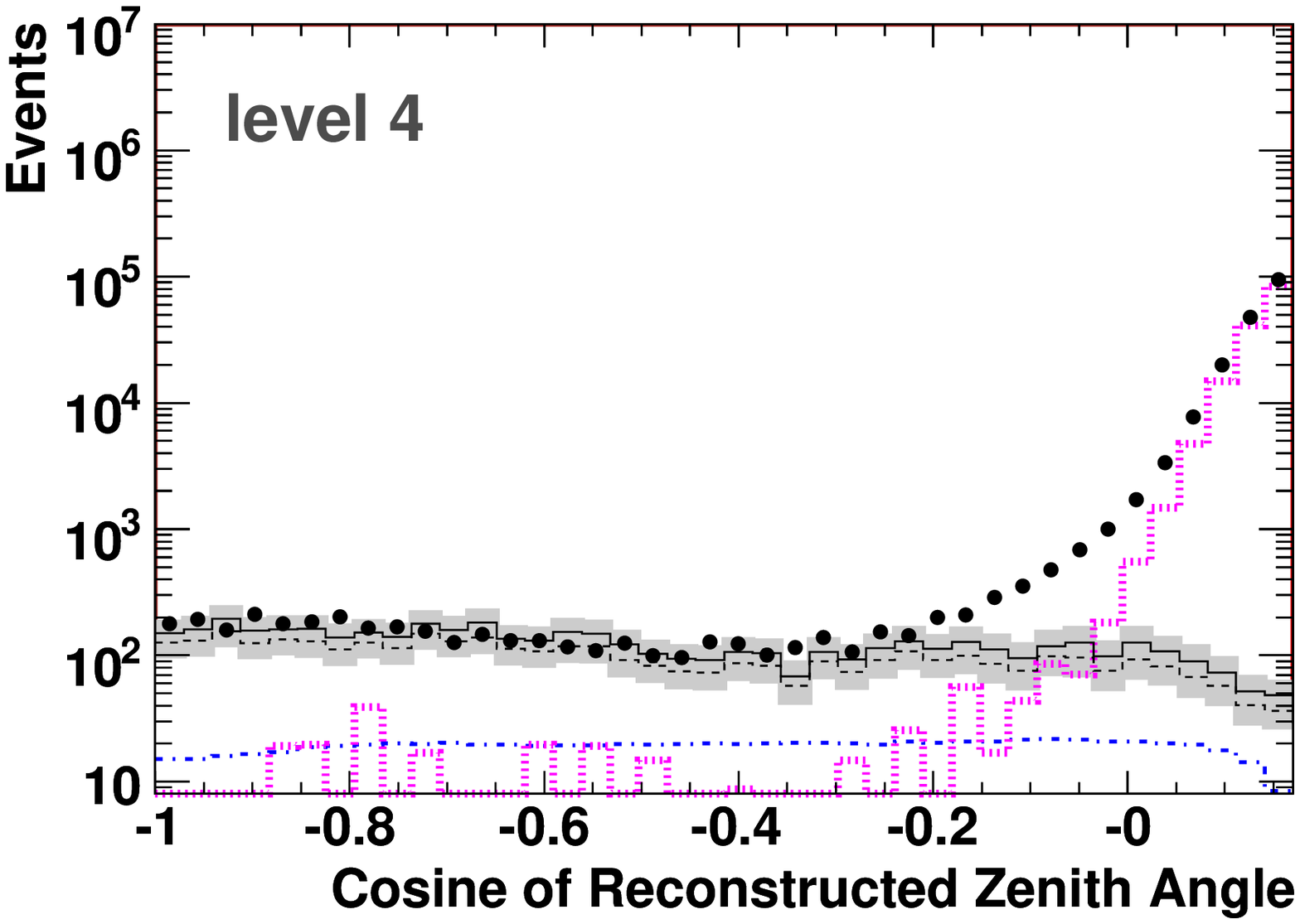}
\includegraphics*[width=0.49\textwidth]{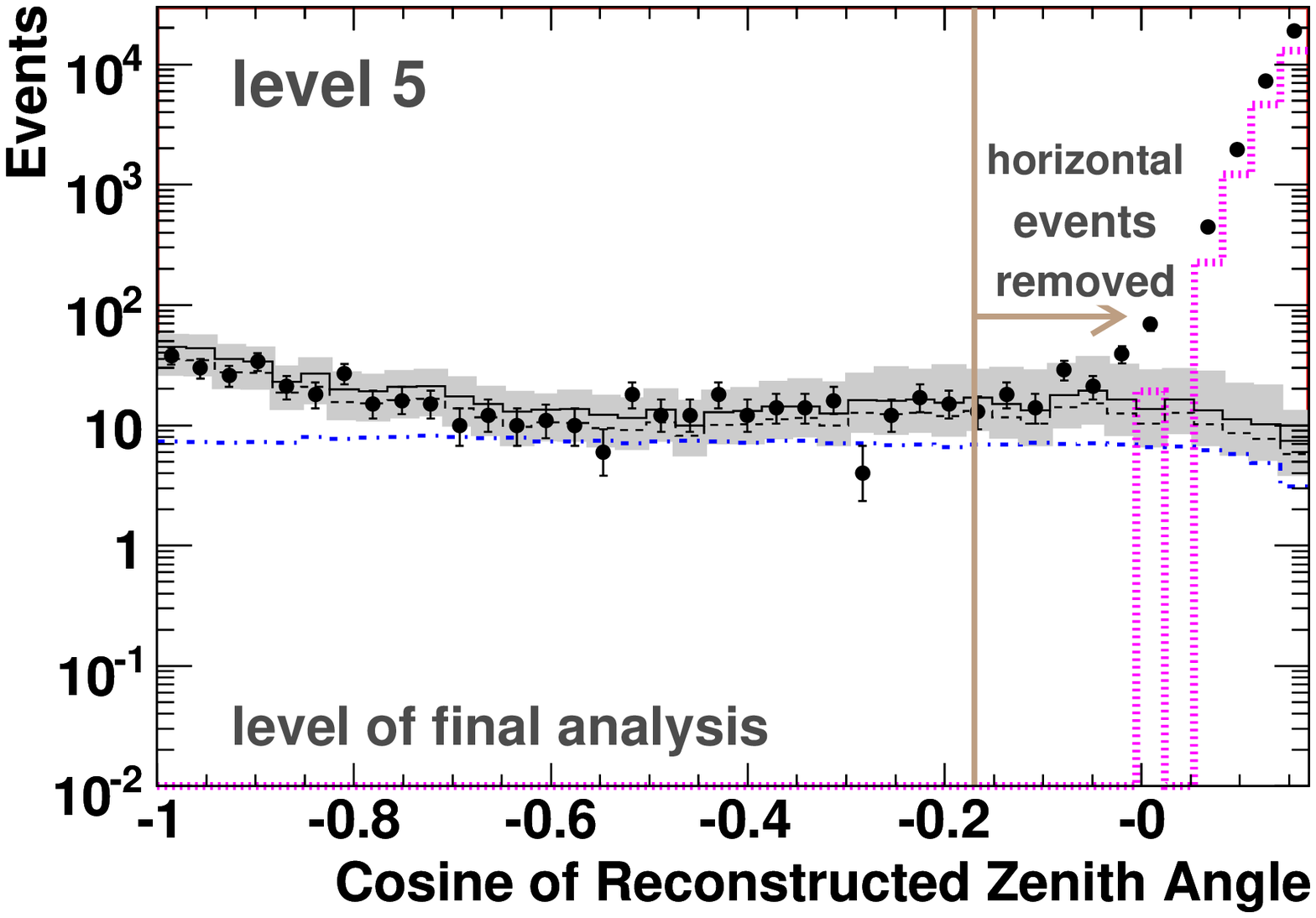}
\caption{\label{zenithplots}The cosine of the zenith angle is plotted for all events surviving
the event quality criteria at a given level. Events at \mbox{cos(zenith) =
$-$1} are traveling straight up through the detector from the Northern
Hemisphere. The initial zenith angle requirement removed events from
0${^o}$ to 80${^o}$ (level 1 - top right). Events reconstructed just above
the horizon appear at the right side of each plot. Each level represents an
increasingly tighter set of quality requirements. As the quality level
increased, misreconstructed downgoing muons were eliminated. To ensure a
clean upgoing sample, events coming from the horizon were discarded by
requiring reconstruction angles greater than 100${^o}$. The final analysis
was performed at level 5 (bottom right) with horizontal events removed.}
\end{figure*}

The event selection requirements were successively tightened, based on the
reconstructed track parameters, establishing five quality levels. At Level
5, used for the final stages of the analysis, the event sample is expected
to contain only truly upgoing events. The zenith angle distribution for the
events at each quality level is shown in Figure \ref{zenithplots}. Although
the entire upgoing zenith angle region is being studied, the event
selection requirements preferentially retain vertically upgoing
events. Horizontal and vertical events must pass the same requirements for
track length and number of direct hits, however this is more difficult for
horizontal events since the detector is not as wide as it is tall.

\begin{figure*}
\centering
\includegraphics*[width=3.4in]{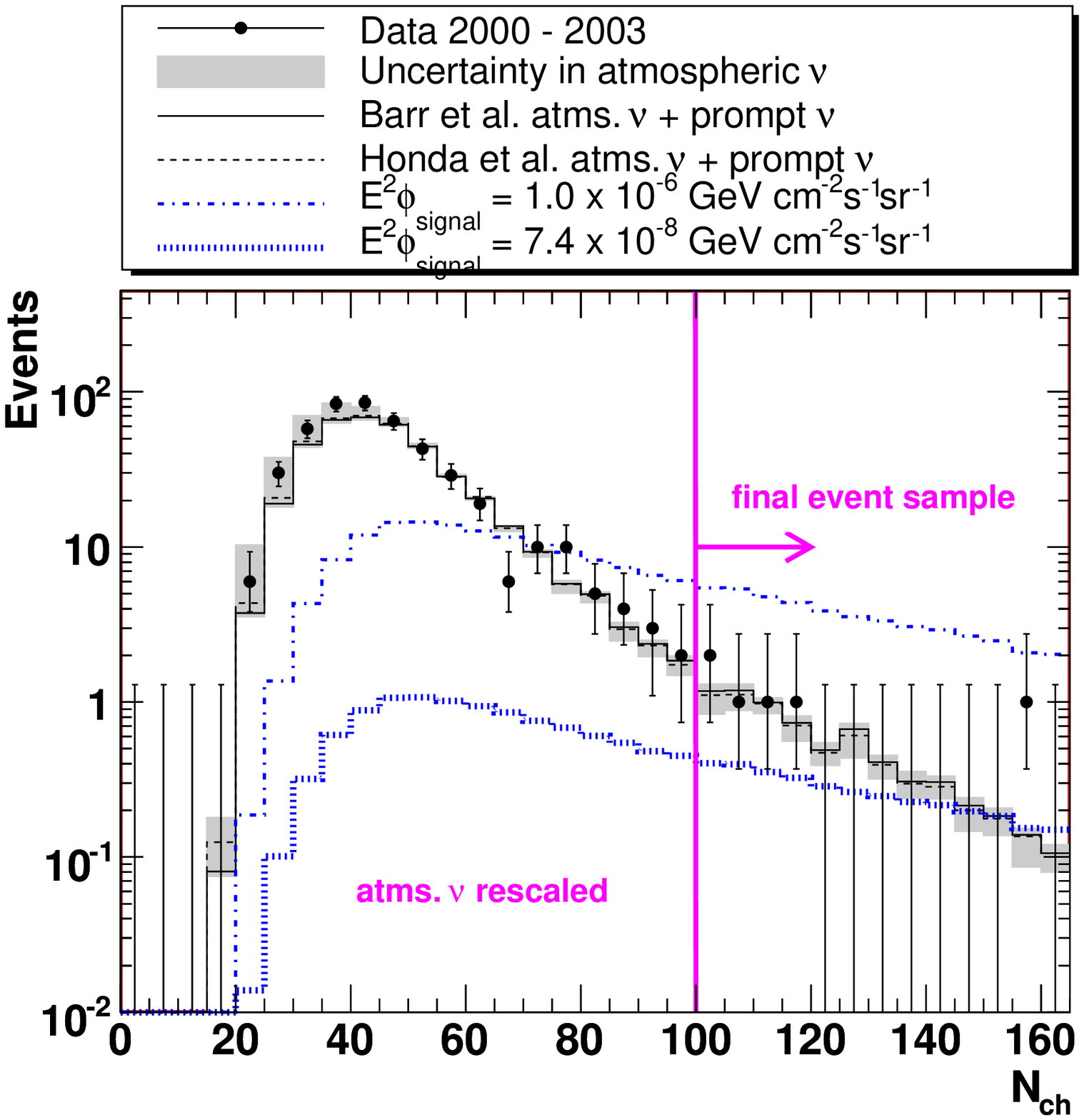}
\caption{\label{nch_sigrescaled}$N_\mathrm{ch}$, or number of OMs
hit. Prediction for both conventional and prompt atmospheric neutrinos are
shown and their uncertainties are represented by the gray band. The central
prompt neutrino flux used here is the average of the Martin GBW
\cite{martin_gbw} and Naumov RQPM \cite{naumov_rqpm_a,naumov_rqpm_b}
models. All atmospheric neutrinos are normalized to the number of data
events in the range 50
\textless $N_\mathrm{ch}$ \textless 100. The lower signal flux curve
corresponds to the limit obtained in this paper.}
\end{figure*}

\subsection{Separating Atmospheric Neutrinos from Astrophysical Neutrinos}

Figure \ref{nch_sigrescaled} shows the $N_\mathrm{ch}$ distribution for
events at Level 5. The optimal place for the energy-correlated event
observable requirement was established with the simulation by minimizing
the expected Model Rejection Factor (MRF) \cite{mrp}. The Feldman-Cousins
method was used to calculate the median upper limit \cite{feldcous}. The
MRF is the median upper limit divided by the number of predicted signal
events for the $\nu_\mu$ signal being tested. The MRF was calculated for
every possible $N_\mathrm{ch}$ value and was at its minimum at
$N_\mathrm{ch} \geq 100$. Hence, the optimal separation of astrophysical
and atmospheric neutrinos is achieved with this $N_\mathrm{ch}$
requirement.

\begin{figure*}
\centering
\includegraphics*[width=3.4in]{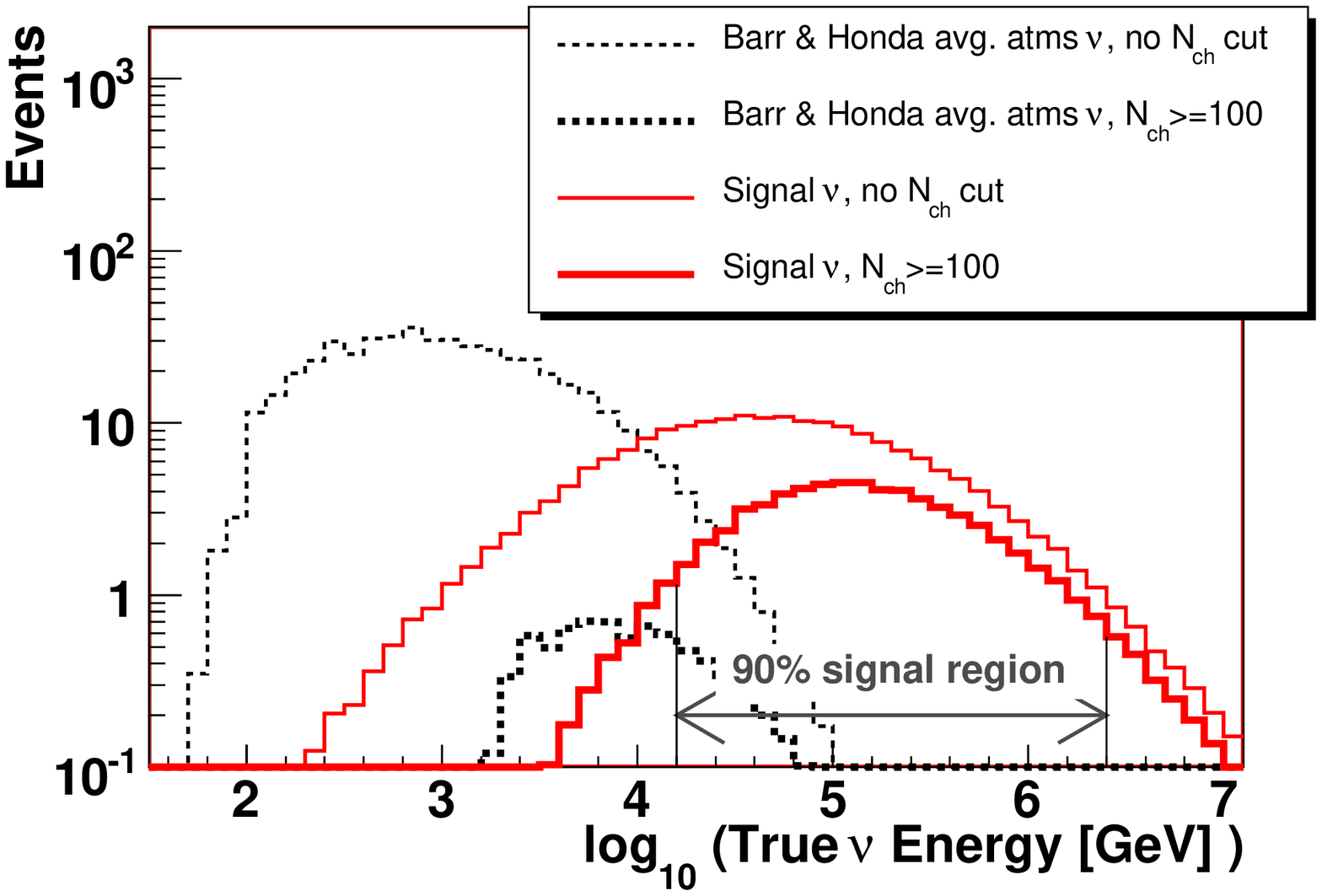}
\caption{\label{trueenplot}The true energy of the simulation is shown for atmospheric
neutrino and signal events. The thin dashed (atms. $\nu$) and solid (signal
$\nu$) curves represent the number of events before the $N_\mathrm{ch} \geq
100$ requirement. The thick dashed and solid lines represent only the
events in the high energy sample.}
\end{figure*}

The final event sample was composed of events which pass all event
selection requirements (Level 5) and have \mbox{$N_\mathrm{ch} \geq 100
$}. After the high $N_\mathrm{ch}$ requirement, the atmospheric neutrino
simulation peaked at 10 TeV, while the signal simulation peaked around 100
TeV (Figure \ref{trueenplot}). The energy range defined by the central 90\%
of the signal with \mbox{$N_\mathrm{ch} \geq 100$} is the energy range for
the sensitivity or limit. For this search, the central 90\% signal region
extends from \mbox{16 TeV} to \mbox{2.5 PeV}.

The efficiency of the detector for neutrinos is quantified by the effective
area. In the energy range relevant to this analysis, it increases with
energy and is further enhanced by including uncontained events. The
effective area is described by the following equation where $N$ represents
the number of observed events and $T$ is the detector livetime:

\begin{equation}
\mbox{$\frac{N}{T} = \int A_\mathrm{eff}^{\nu}(E_{\nu},\Omega) \Phi_{\nu}
d\Omega dE$}~. 
\label{effareaequation}
\end{equation}

The effective area as a function of energy is shown for different zenith
angle regions in Figure \ref{effareaplots} (and is tabulated in Appendix
B). At energies greater than 10$^{5}$ GeV, the Earth begins to be opaque to
neutrinos depending on direction and the highest energy events are most
likely to come from the region around the horizon \cite{lisa_icrc2005}. In
Figure \ref{effareaplots}, the effective area decreases at high energy
because tracks with zenith angles between 80$^{\circ}$ and 100$^{\circ}$
were discarded. Most of the events that were removed were high energy
events from the horizon.

\begin{figure*}
\centering
\includegraphics*[width=3.4in]{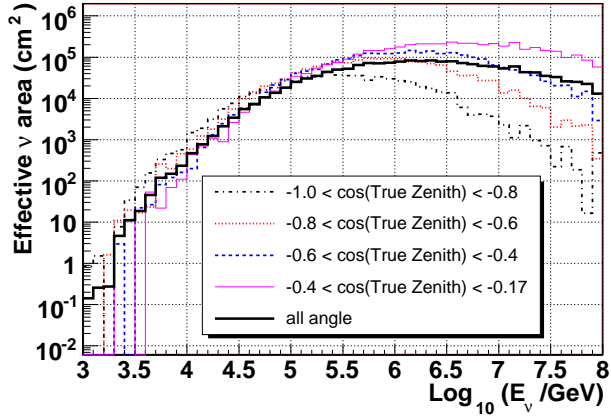}
\caption{\label{effareaplots}Effective area for $\nu_{\mu}$ as a function
of the true simulated energy at the Earth's surface in intervals of cosine
of the true zenith angle. The effective area is the equivalent area over
which the detector would be 100\% efficient for detecting neutrinos. The
absorption of neutrinos and reduction of their energy via neutral current
interactions in the Earth are taken into account. The angle-averaged
effective area is represented by the solid black line.}
\end{figure*}

\section{\label{section_systematics}Systematic Uncertainties}

A discovery is made if an excess of events over the predicted background is
observed in the data. However, due to uncertainties in the simulation, the
number of signal and background events predicted may not accurately reflect
the true signal and background. Theoretical uncertainties exist in the
atmospheric neutrino flux models for several reasons. The cosmic ray
spectrum is very uncertain at high energy and hadronic interactions in this
energy range are not well understood. There are also detector-related
uncertainties. Photons scatter more in dirty or bubble-laden ice. Hence,
our incomplete understanding of the dust layers in the ice and the bubbles
in the hole ice (formed from water that refroze after deployment of the
OMs) add uncertainty to our models \cite{icepaper}. There are also
uncertainties in the simulation associated with the modeling of light
propagation in the ice and with the optical module sensitivity. These
contributions are considered individually to see how they affect the number
of simulated events in the final sample. The number of experimental data
events remaining after the final energy requirement \mbox{($N_\mathrm{ch}
\geq$ 100)} is then compared to the range of predicted background and
signal events when uncertainties are considered.

\subsection{\label{subsection_theoretical_unc}Theoretical Uncertainty in the Background}

For this analysis, two models based on the work of Barr \textit{et al.}
\cite{bartol2004,barronlinetables,icrc2001gaisser} and Honda \textit{et al.} \cite{honda2004} were
considered equally likely options for the background atmospheric neutrino
simulation. These two models are recent calculations that cover the highest
and lowest portion of the atmospheric neutrino flux band created by
uncertainties in the primary cosmic ray flux and the high energy hadronic
interaction models. Since these models do not extend to the high energies
needed for this analysis, the models were extrapolated to higher energies
based on the procedure described in Appendix C. Differences between the
Barr \textit{et al.} and Honda \textit{et al.} models are also summarized
there.

Conventional atmospheric neutrinos from the decay of pions and kaons are
not the only source of atmospheric background. Above 50 TeV - 1 PeV, the
source of atmospheric neutrinos is expected to change
\cite{zhv_charm,naumov_rqpm_a,naumov_rqpm_b,martin_gbw,prompt_lepton_cookbook}. Semileptonic decays of short-lived charmed
particles become the main contributor to the atmospheric neutrino
flux. Since these charmed particles decay quickly before they can lose much
energy, the resulting neutrinos are called \textit{prompt} neutrinos. At
these energies, charm quarks are produced primarily via gluon-gluon
fusion. Uncertainties in the gluon distribution at low Bjorken $x$ lead to
uncertainties in this prompt lepton flux.

Uncertainties were included for both conventional atmospheric neutrino
models. The uncertainty in the cosmic ray spectrum was estimated as a
function of energy based on the spread of values measured by many cosmic
ray experiments \cite{gaisser_honda_review}. These uncertainties were added
in quadrature with the estimated uncertainty due to choosing different
hadronic interaction models
\cite{bartol2004,honda2004,gaisser2005}. Uncertainties were also estimated
based on the spread of predictions surrounding the unknown prompt neutrino
flux. Unless mentioned otherwise, when prompt neutrinos were included in
this work, the average of the Martin GBW (Golec-Biernat and W\"usthoff)
\cite{martin_gbw} and Naumov RQPM (Recombination Quark Parton Model)
\cite{naumov_rqpm_a,naumov_rqpm_b} models is shown. This is henceforth
called the central prompt neutrino model.

All of the uncertainty factors for the total (conventional + prompt)
atmospheric neutrino flux were combined and are shown as a function of
energy in Appendix C. Since the true energy of every simulated event is
known, each event was given a weight based on the maximum uncertainty
estimated for that neutrino energy. As a result, three predictions for the
number of atmospheric neutrinos in the final high energy sample were made
(the model, the model plus maximum energy-dependent uncertainty, the model
minus maximum energy-dependent uncertainty). Since both the Barr \emph{et
al.} and Honda \emph{et al.} fluxes were considered equally likely, the
central prompt neutrino flux was added to both predictions. Then
uncertainties were added and subtracted to both of these total atmospheric
neutrino fluxes, creating six different background possibilities.

\subsubsection{\label{section_constrainmodels}Normalizing the Atmospheric
Neutrino Simulation to the Data}

After all but the $N_\mathrm{ch}$ event selection requirements were
fulfilled, the $N_\mathrm{ch}$ distribution for the observed low energy
events was inconsistent with that for the total atmospheric neutrino
simulation in normalization. Each of the six atmospheric neutrino
background predictions was renormalized to match the number of data events
observed in the low $N_\mathrm{ch}$ region, where the signal was
insignificant compared to the background. By rescaling the simulation to
the number of observed data events, the uncertainty in the atmospheric
neutrino flux was reduced to the uncertainty in the spectral shape.

Since some of the atmospheric neutrino models predicted more events than
the data while others predicted less, renormalization of the models to the
data brought the simulated models into closer agreement. The
renormalization is explained in greater detail in Appendix~C.

Since the purpose of this normalization was to correct for theoretical
uncertainties in the atmospheric neutrino background prediction, it was not
applied to the simulated neutrino signal.

\subsection{\label{subsection_simulation_uncertainties}Simulation Uncertainties}

To assure that the detector response to high energy events
(\mbox{$N_\mathrm{ch} \geq 100$}) is understood, it is important to study
energetic events while simultaneously keeping the high energy upgoing
events blind to the analyzer. To this end, an inverted analysis was
performed in which high quality downgoing tracks were selected from the
initial data set. The advantage of studying high quality downgoing tracks
is that large data sets are available to study both the high and low
energy events. When the data and simulation observable distributions are
not perfectly matched, imposing event quality requirements may result in
removing different fractions of the simulation in comparison with the
data. The inverted analysis was used to study this systematic effect.

\subsubsection{\label{subsection_inverted_analysis}Inverted Analysis Using
Atmospheric Muons}

\begin{figure*}
\centering
\includegraphics*[width=3.4in]{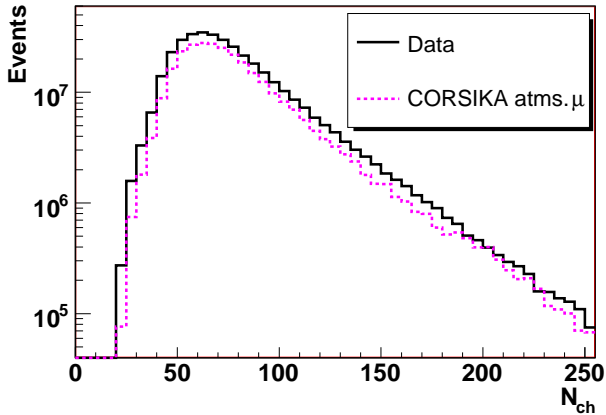}
\caption{\label{invertednch}In the inverted analysis, the highest quality
downgoing events were studied. The $N_\mathrm{ch}$ distribution is shown for all
events which survive the inverted quality requirements.}
\end{figure*}

For the inverted analysis, all event quality requirements described
previously (and summarized in Appendix A) were applied, but events were
selected based on a high probability of being downgoing rather than upgoing
tracks.

When compared to the downgoing experimental data, small shifts were
observed in the peaks of the simulated distributions for the number of
direct hits ($N_\mathrm{dir}$), the smooth distribution of hits along the
track ($S$), the event-by-event track resolution ($MR$) and likelihood of
being downgoing muon tracks rather than upgoing (Inverted Likelihood Ratio,
$ILR$). These discrepancies are most likely due to inaccurate modeling of
optical ice properties in the simulation, since it is technically
challenging to implement a detailed description of photon propagation
through layered ice.

If multiple parameters are correlated, it is possible that mismatches in
one parameter may affect the agreement between data and simulation in
another. In order to study these effects, the differences in the data and
simulation were analyzed at the level where no quality criteria had been
applied. The simulated distributions needed to be shifted to larger values
by approximately 10\% for $N_\mathrm{dir}$, 8\% for $S$, 5\% for $MR$ and
1\% for $ILR$. When simultaneous corrections to the simulation for all of
these effects were applied, the downgoing data and simulation were in
better agreement for all parameter distributions. Later in the analysis,
these shifts were applied to the upgoing analysis. The number of background
and signal events appearing in the final upgoing sample was recalculated
based on these simulation shifts.

\subsubsection{\label{subsubsection_det_response}Uncertainties in Detector Response}

The downgoing sample from the inverted analysis was also used to study how
well the detector response was simulated in the high energy
(\mbox{$N_\mathrm{ch} \geq 100$}) regime. Using downgoing data and
atmospheric muon simulation, a ratio of the number of events was taken as a
function of $N_\mathrm{ch}$ from the histograms shown in Figure
\ref{invertednch}. If the simulation perfectly described the data, the
shapes of the $N_\mathrm{ch}$ distributions would match and this ratio
would be flat. The downgoing ratio was mostly flat, but slightly increased
at large $N_\mathrm{ch}$ where low statistics introduced large
uncertainties. The statistical uncertainty aside, a scenario was considered
in which the downgoing data to simulation ratio truly increased as
$N_\mathrm{ch}$ increased. Under this scenario, the simulation is
renormalized by a larger factor at high $N_\mathrm{ch}$ to replicate the
data. This $N_\mathrm{ch}$-dependent renormalization was then applied to
the upgoing simulation used for the main part of the analysis. This
non-linear normalization factor had a negligible effect in the number of
atmospheric neutrinos predicted in the final sample of events with
\mbox{$N_\mathrm{ch} \geq 100$}. However, the high energy signal simulation
event rate increased by 25\% when this non-linear $N_\mathrm{ch}$ effect
was included. This uncertainty was incorporated in the final limit
calculation that will be described in the next section.

Detection efficiency also depends on the OM sensitivity. This parameter of
the simulation was modified and new simulated events were generated. After
comparing the data and simulation with different OM sensitivities, a 10\%
uncertainty in the total number of events due to inaccurate modelling of
the OM detection sensitivity was incorporated into the final upper limit
calculation.

The systematic errors due to the neutrino interaction cross-section, rock
density (below the detector), and muon energy loss do not contribute
significantly to this analysis \cite{pointsource5yr}.

\section{\label{section_relateupdown}Relationship between Upgoing and Downgoing Events}

In addition to using the inverted analysis to study high energy events and
the bias introduced by inaccurate simulation, the downgoing events can be
used as a calibration beam for the upgoing atmospheric neutrino flux. To do
this, the same simulation package (CORSIKA v6.030, QGSJET01, $\Phi_{primary}
\propto$ E$^{-2.7}$) was used to describe the downgoing
atmospheric muons and the upgoing atmospheric neutrinos
\cite{dcorsikaneutrinos}.

As shown in Table \ref{invertedtable}, the ratio of experimental data to
CORSIKA downgoing muon simulation was relatively constant as the event
selection became more discriminating. The simulation does not match the
data normalization and this may be a consequence of the theoretical
imperfections in the CORSIKA simulation (mainly due to the hadronic
interaction model (QGSJET01) and uncertainty in the primary spectrum ($\Phi
\propto$ E$^{-2.7}$)). Another contributing factor to the normalization
difference may be that light propagation in the layered ice is modeled
inaccurately.  When the upgoing CORSIKA atmospheric neutrinos are rescaled
based on the downgoing muons, then the upgoing experimental data and
CORSIKA atmospheric neutrino simulation are in good agreement for the
number of low energy events in the final sample. This can only be seen when
the tightest criteria are applied because misreconstructed muons and
coincident muons contaminate the data sample when the quality requirements
are loose. For instance, at level 5 in the inverted analysis, the ratio of
downgoing data to simulation was 1.22. For the upgoing analysis at level 5,
146 events were observed and 124.9 CORSIKA atmospheric neutrinos were
predicted. When adjusted based on the inverted analysis, \mbox{152.4 ($ =
124.9\times 1.22$)} CORSIKA atmospheric neutrinos were predicted, which is
in good agreement with the observed value. This shows that it is possible
to adjust the normalization of the upgoing events based on the downgoing
observations (when the up and downgoing simulation use the same input
assumptions about the spectrum and interaction model).

\renewcommand{\arraystretch}{0.65}
\begin{table*}
\small
\begin{tabular} {|l|l|l|l|l|l|}
\hline
 & L1 & L2 & L3 & L4 & L5$^{*}$ \\
\hline
\normalsize
\textbf{\textit{Downgoing}} & & & & & \\
\hline
\small
data ($\times 10^{8}$) & 7.88 & 6.70 & 6.05 & 5.89 & 2.59 \\
\hline
CORSIKA & & & & & \\
atms. $\mu$($\times 10^{8}$) & 6.63 & 5.75 & 5.12 & 5.01 &
2.12 \\
\hline
ratio & 1.19 & 1.17 & 1.18 & 1.18 & 1.22 \\
\hline
\normalsize
\textbf{\textit{Upgoing}} & & & & & \\
\hline
\small
signal & 325 & 241 & 191 & 185 & 103 \\
\hline
coinc $\mu$ & 2570 & 268 & 45.8 & 29.4 & 0 \\
\hline
misreconstructed & & & & & \\
CORSIKA & & & & & \\
atms. $\mu$ & 37800 & 2570 & 148 & 34.2 & 0 \\
\hline
Barr \textit{et al.} & & & & & \\
atms. $\nu$ & 681 & 526 & 393 & 380 & 194 \\
\hline
Honda \textit{et al.} & & & & & \\
atms. $\nu$ & 513 & 400 & 300 & 290 & 149 \\
\hline
Martin GBW prompt $\nu$ & 1.9 & 1.9 & 1.6 & 1.5 & 0.7 \\
\hline
Naumov RQPM prompt $\nu$ & 18.9 & 18.9 & 16.0 & 15.5 & 7.5 \\
\hline
CORSIKA & & & & & \\
atms. $\nu$ & 440 & 335 & 251 & 243 & 125 \\
\hline
Adjusted & & & & & \\
CORSIKA & & & & & \\
atms. $\nu$ & 524 & 392 & 296 & 286 & 152 \\
\hline
data & 276894 & 24422 & 1269 & 531 & 146 \\
\hline
\multicolumn{6}{r}{$^{*}$L5 = level of final analysis}\\
\end{tabular}
\caption{\label{invertedtable}The number of low energy events (\mbox{50
\textless $N_\mathrm{ch}$ \textless 100}) at a given level (see Appendix A)
for the different types of simulation and experimental data. The top
portion of the table presents results from the inverted analysis. The main
upgoing analysis is summarized in the lower portion of the table. Note that
the upgoing data and adjusted CORSIKA atmospheric neutrino flux are in good
agreement when the CORSIKA neutrino events are adjusted by the scale factor
determined in the downgoing analysis. This agreement can be seen at the
tightest quality levels because all misreconstructed backgrounds have been
removed.}
\end{table*}
\renewcommand{\arraystretch}{1}

\begin{table}
\footnotesize
\begin{tabular} {|l|c|}
\hline
\multicolumn{2}{|l|}{\textbf{Systematic Uncertainty}}\\
\hline
Theoretical uncertainty in atms. $\nu$ flux & See Figure
\ref{energyuncertainty} \\
\hline
Number of Direct Hits & 10\% \\
\hline
Smoothness & 8\% \\
\hline
Median Resolution & 5\% \\
\hline
Inverted Likelihood Ratio & 1\%  \\
\hline
\textbf{Total background uncertainty} & \textbf{+19\% / -18\%} \\ 
\hline
\hline
Non-linear detector response at high $N_\mathrm{ch}$ & 25\%  \\
\hline
OM sensitivity & 10\%  \\
\hline
\textbf{Total signal efficiency uncertainty} &
\textbf{+/- 27\%}\\
\hline
\end{tabular}
\caption{\label{totaluncertainty} The systematic error was estimated with
several techniques. The theoretical uncertainty in the atmospheric neutrino
flux was estimated as a function of energy (Section
\ref{subsection_theoretical_unc}). Using the inverted analysis, shifts were
observed between the data and simulation in four parameters (Section
\ref{subsection_inverted_analysis}). When each of the above mentioned uncertainty factors was applied to
the atmospheric neutrino simulation, the resulting spread in the number of
events predicted in the $N_\mathrm{ch} \geq 100$ sample indicated that the
total background uncertainty was +19\% / -18\% of the average predicted
background, 7.0 events. The non-linear response of the detector in
$N_\mathrm{ch}$ was estimated as 25\% (Section
\ref{subsubsection_det_response}). When added in quadrature with the 10\%
uncertainty in OM sensitivity (Section \ref{subsubsection_det_response}),
the total signal efficiency uncertainty was +/-27\%.}
\end{table}

\section{\label{section_results}Results}

We calculated a confidence interval based on the number of events in the
final $N_\mathrm{ch} \geq 100$ sample of the predicted background and
signal and the observed data. Statistical and systematic uncertainties were
incorporated into the confidence interval such that the true, but unknown,
value of the diffuse flux of astrophysical neutrinos is contained within
the interval in 90\% of repeated experiments. A hybrid frequentist-Bayesian
method based on the work of Cousins and Highland \cite{cousinshighland} was
used to construct a confidence belt with systematic uncertainties. The
likelihood ratio ordering was based on the unified confidence intervals
explained by Feldman and Cousins \cite{feldcous}. The uncertainty in the
detection efficiency of the signal was set at 27\% (10\% for optical module
sensitivity added in quadrature with 25\% for non-linearity in the
$N_\mathrm{ch}$ spectrum when data and simulation are compared). Systematic
uncertainties on the number of background events in the final sample were
also included in the confidence belt construction. Inclusion of the signal
and background uncertainties followed the methods described by Conrad
\emph{et al.} \cite{conrad} and Hill \cite{hillci}.

In constructing the flat Bayesian prior for the background, twelve
atmospheric neutrino models were considered equally likely. The twelve
predictions were derived as follows. Initially, two background predictions
were considered, Barr \emph{et al.} and Honda \emph{et al.}, each with the
central prompt neutrino flux added. To include systematic uncertainties in
the models, maximum uncertainties were added and subtracted from each model
(Section \ref{subsection_theoretical_unc}). Hence, the six predictions were
named Barr \emph{et al.} maximum, nominal and minimum and Honda
\emph{et al.} maximum, nominal and minimum. The number of events predicted
for the background in the final sample is listed in Appendix C. To account
for systematic uncertainties in the detector response, the simulation was
shifted in four different parameters as described in Section
\ref{subsection_inverted_analysis}. This simulation shift was performed on
each of the 6 models described above, hence creating a total of 12
different atmospheric neutrino predictions that were used in the confidence
belt construction. The number of events predicted by the 6 models with
shifted simulation was within 10\% of each number reported in Appendix C.

\subsection{\label{subsection_e2results}Results for \mbox{$\Phi \propto$ E$^{-2}$}}

The signal hypothesis consisted of a flux \mbox{E$^{2} \Phi$ = 1.0 $\times$
10$^{-6}$ GeV cm$^{-2}$ s$^{-1}$ sr$^{-1}$}. At this signal strength, 66.7
signal events were expected in the final \mbox{$N_\mathrm{ch} \geq 100$}
data. (This value assumes half of the correction from the simulation shifts
since 68.4 events were predicted in the final selection, but the number of
events decreased to 65.0 when the simulation shifts were applied.) The
sensitivity was obtained from the slice of the confidence belt
corresponding to zero signal strength. The median observation assuming no
signal was seven events, giving a median event upper limit of 6.36 and hence
a sensitivity of \mbox{9.5 $\times$ 10$^{-8}$ GeV cm$^{-2}$ s$^{-1}$
sr$^{-1}$}.

\begin{figure*}

\centering

\includegraphics*[width=0.99\textwidth]{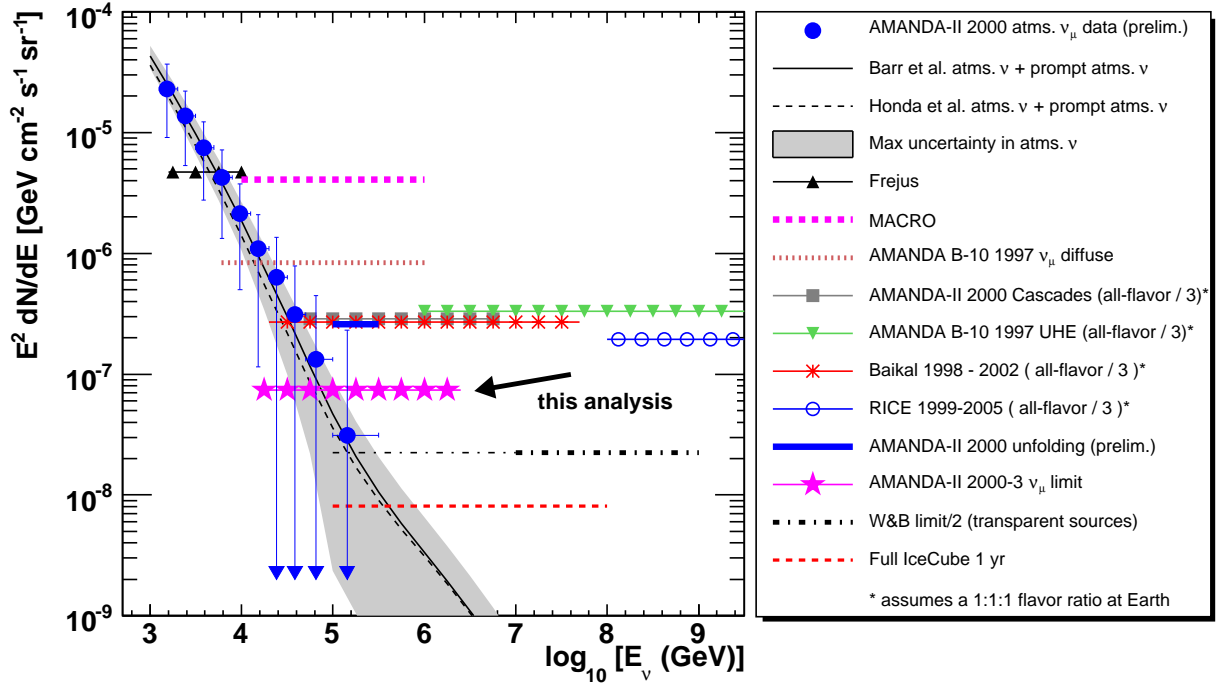}
\caption{\label{limitplot_allflavor} The upper limits on the $\nu_{\mu}$ flux from sources with an
$E^{-2}$ energy spectrum are shown for single and all-flavor
analyses. All-flavor upper limits have been divided by three, assuming that
the neutrino flavor ratio is 1:1:1 at Earth. The Fr\'ejus \cite{frejus},
MACRO \cite{macro}, and AMANDA-B10 \cite{diffuse97} upper limits on the
$\nu_{\mu}$ flux are shown, as well as the unfolded atmospheric spectrum
from 2000 \mbox{AMANDA-II} data \cite{kirsten_unfolding}. The \mbox{AMANDA-II} all-flavor limit from
2000 \cite{cascades2000}, the AMANDA-B10 UHE limit \cite{uhe1997}, the Baikal five year limit
\cite{baikal} and the RICE six year limit \cite{rice} have all been
adjusted for the single flavor plot. The $\Phi \propto$ E$^{-2}$ limit from
this analysis is a factor of four above the Waxman-Bahcall upper
bound. Although not shown, this analysis excludes the $\Phi \propto$
E$^{-2}$ predictions made by Nellen, Mannheim and Biermann
\cite{nellen} and Becker, Biermann and Rhode \cite{becker} and constrains
the MPR upper bound for optically thick pion photoproduction sources
\cite{mpr}. The IceCube sensitivity for a full detector was estimated with
AMANDA software
\cite{icecubesensitivity}.}

\end{figure*}

When the data with \mbox{$N_\mathrm{ch} \geq 100$} were revealed, six data
events were observed. This was consistent with the average expected
atmospheric neutrino background of 7.0 events (after averaging all models
that have been rescaled to the low energy data). Information about the observable
quantities for the final six events can be seen in Table
\ref{thefinalsixtable}. The final $N_\mathrm{ch}$ distribution is shown in Figure
\ref{nch_sigrescaled}. The total number of events predicted for the signal
and background can be compared to the observed data in Table
\ref{invertedtable_less100} (\mbox{$N_\mathrm{ch}$ \textless 100}) and in Table
\ref{invertedtable_more100} (\mbox{$N_\mathrm{ch} \geq 100$}). With
uncertainties included, the upper limit on a diffuse \mbox{$\Phi \propto$
E$^{-2}$} flux of muon neutrinos at Earth (90\% confidence level) with the
\mbox{AMANDA-II} detector for 2000 -- 2003 is
\mbox{7.4 $\times$ 10$^{-8}$ GeV cm$^{-2}$ s$^{-1}$ sr$^{-1}$} for \mbox{16
TeV} to \mbox{2.5 PeV}. The results are compared to other neutrino limits
in Figure \ref{limitplot_allflavor}.

\renewcommand{\arraystretch}{0.65}
\begin{table*}
\footnotesize
\begin{tabular} {|r|r|r|r|r|r|r|r|}
\hline
Event & \textbf{1} & \textbf{2} & \textbf{3} & \textbf{4} & \textbf{5} & \textbf{6} & \textit{required}\\
 & & & & & & & \textit{value}\\ 
\hline
Year & 2001 & 2001 & 2001 & 2001 & 2002 & 2003 &  \\
\hline
Day of Year & 118 & 186 & 210 & 274 & 226 & 182 & \\
\hline
$N_\mathrm{ch}$ & 102 & 106 & 157 & 116 & 100 & 111 & $\geq$100 \\
\hline
Track Length [m] & 206.7 & 221.8 & 197.7 & 178.2 & 180.4 & 207.6 &
\textgreater 170\\
\hline
Number of & & & & & & & \\
Direct Hits & 27 & 32 & 30 & 22 & 29 & 29 & \textgreater 13 \\
\hline
Zenith Angle [$^{\circ}$] & 107.3 & 121.6 & 106.1 & 101.8 & 123.8 & 113.3 &
\textgreater 100 \\
\hline
Median Resolution [$^{\circ}$] & 2.4 & 1.4 & 1.8 & 3.0 & 1.6 & 2.8 & \textless
4.0 \\
\hline
\end{tabular}
\caption{\label{thefinalsixtable}Observable and reconstructed qualities are shown for the final six
events. In addition, events fulfilled requirements based on the
reconstructed values of their smoothness ($S$) and their upgoing
vs. downgoing likelihood ratios.}
\end{table*}
\renewcommand{\arraystretch}{1}

\renewcommand{\arraystretch}{0.65}
\begin{table*}
\small
\begin{tabular} {|r|r|r|}
\hline
\textbf{Experiment} & \textbf{Upper Limit} & \textbf{Energy Range} \\
 & [GeV cm$^{-2}$ s$^{-1}$ sr$^{-1}$] & log$_{10}$ [E$_{\nu}$ (GeV)] \\
\hline
\multicolumn{3}{|c|} {\textit{Muon neutrinos only}} \\
\hline
Fr\'ejus \cite{frejus} & $5.0 \times 10^{-6}$ & $\sim$3.4 \\
\hline
MACRO \cite{macro} & $4.1 \pm 0.4 \times 10^{-6}$ & 4.0 -- 6.0\\
\hline
AMANDA-B10 \cite{diffuse97} & $8.4 \times 10^{-7}$ & 3.8 -- 6.0 \\ 
\hline
AMANDA-II (this analysis) & $7.4 \times 10^{-8}$ & 4.2 -- 6.4 \\
\hline
\multicolumn{3}{|c|} {\textit{All neutrino flavors}} \\
\hline
Baikal \cite{baikal} & $8.1 \times 10^{-7}$ & 4.3 -- 7.7 \\
\hline
AMANDA-B10 \cite{uhe1997} & $0.99 \times 10^{-6}$ & 6.0 -- 9.5 \\
\hline
AMANDA-II \cite{cascades2000} & $8.6 \times 10^{-7}$ & 4.7 -- 6.7 \\
\hline
\end{tabular}
\caption{\label{allexperimentresults} Upper limits for the diffuse flux of
extraterrestrial neutrinos as reported by a number of experiments. The
first four analyses only constrain the flux of $\nu_{\mu} +
\bar{\nu}_{\mu}$ , while the last three constrain the total neutrino flux,
($\nu_{e} + \bar{\nu}_{e} + \nu_{\mu} + \bar{\nu}_{\mu} + \nu_{\tau} +
\bar{\nu}_{\tau}$).}
\end{table*}
\renewcommand{\arraystretch}{1}

\subsection{\label{subsection_diffspectra}Results for Other Energy Spectra}

Other signal models were also tested with this data set. Due to their
different energy spectra, the $N_\mathrm{ch}$ requirement was reoptimized
by minimizing the MRF with each signal model. For signal models with softer
spectra than \mbox{$\Phi \propto$ E$^{-2}$}, a lower $N_\mathrm{ch}$
requirement was optimal, \mbox{$N_\mathrm{ch} \geq 71$}. Four prompt
neutrino models \cite{zhv_charm,naumov_rqpm_a,naumov_rqpm_b,martin_gbw} and
one astrophysical neutrino model \cite{loeb_waxman_starburst} were tested
under these conditions. One astrophysical model was optimized at
\mbox{$N_\mathrm{ch} \geq 86$}
\cite{mpr}. Two astrophysical neutrino models with harder spectra than
\mbox{$\Phi \propto$ E$^{-2}$} were tested with a higher energy
requirement, \mbox{$N_\mathrm{ch} \geq 139$} \cite{mpr,sdss,sdss_revision}.

Results of these searches are summarized in Table
\ref{othermodelstable}. The normalization of the overall number of low
energy atmospheric neutrinos to data was performed over the region
\mbox{$50 < N_\mathrm{ch} < 100$} for the harder spectra
\mbox{($N_\mathrm{ch} \geq 139$)}, and over \mbox{$50 < N_\mathrm{ch} <
71$} and \mbox{$50 < N_\mathrm{ch} < 86$} for the softer spectra.

When the data from the \mbox{$N_\mathrm{ch} \geq 139$} region were examined,
there was good agreement with the expected atmospheric neutrino background
(1 event observed on a backround of 1.55). For \mbox{$N_\mathrm{ch} \geq
86$}, 14 events were observed while an average of 12.9 background events
were predicted. However, 37 events were observed while only 27.4 events
were expected for \mbox{$N_\mathrm{ch} \geq 71 $}, leading to a two-sided
confidence interval. Since the chance probability of observing 37 or more
events on this background is 4\%, we do not exclude the background-only
null hypothesis. The 90\% confidence interval for $\mu$ is shown for each
model in Table \ref{othermodelstable} and upper limits are calculated based on the upper bound
of each confidence interval. If the MRF is greater than 1, then the model
is not ruled out based on observations from this four-year data set. Since
more events were observed in the data than were predicted by the background
simulation for \mbox{$N_\mathrm{ch} \geq 71$}, the upper limit on those
five models is roughly a factor of three worse than the sensitivity.

\subsubsection{\label{section_astronu}Astrophysical Neutrinos}

The first astrophysical neutrino model tested with the \mbox{$N_\mathrm{ch}
\geq 139$} requirement was initially proposed by Stecker, Done, Salamon and
Sommers \cite{sdss}. The flux tested in this analysis includes the revision
in the erratum of their original paper \cite{sdss} and the factor of 20
reduction by Stecker in 2005 \cite{sdss_revision}. The model predicts a
flux ($\Phi_\mathrm{SDSS}$) of high energy neutrinos from the cores of
AGNs, especially Seyfert galaxies. Based on the present data, the upper
limit on this flux is 1.6$ \cdot \Phi_\mathrm{SDSS}$. The best previous
limit on this model was established by the Baikal experiment, with an upper
limit of 2.5 $ \cdot \Phi_\mathrm{SDSS}$ \cite{baikal}.

Mannheim, Protheroe and Rachen (MPR) \cite{mpr} computed an upper bound for
neutrinos from generic optically thin pion photoproduction sources
($\tau_{n\gamma}<1$), as well as an upper bound for neutrinos from AGN
jets. (In addition, they calculated an upper bound for generic optically
thick ($\tau_{n\gamma}\gg1$) pion photoproduction sources assuming a
\mbox{$\Phi \propto$ E$^{-2}$} spectrum, but this is constrained by the
results discussed in the previous section.) The upper bounds do not
necessarily represent physical neutrino energy spectra, but were
constructed by taking the envelope of the ensemble of predictions for
smaller energy ranges. Each flux prediction within the ensemble was
normalized to the observed cosmic ray proton spectrum.

Nonetheless, the \textit{shapes} of these two upper bounds were tested as
if they were models. However, one should be careful not to misinterpret the
results. A limit on a model implies a change in the normalization of the
entire model. A limit on an upper bound only implies a change in
normalization of the bound in the energy region where the detector energy
response to that spectral shape peaks.

The MPR AGN jet upper bound was tested with the \mbox{$N_\mathrm{ch} \geq
139$} requirement. The upper limit on this spectrum is 2.0$ \cdot
\Phi_\mathrm{MPR AGN}$. In comparison, the Baikal upper limit on this spectrum
is 4.0$ \cdot \Phi_\mathrm{MPR AGN}$.

The MPR upper bound for optically thin sources was tested with a
\mbox{$N_\mathrm{ch} \geq 86$} requirement. The limit on this spectrum and
normalization is 0.22$ \cdot \Phi_\mathrm{MPR \tau < 1}$. 

The remaining neutrino searches were conducted with the lower $N_\mathrm{ch}$
requirement, \mbox{$N_\mathrm{ch} \geq 71$}. A signal hypothesis involving
neutrinos from starburst galaxies \cite{loeb_waxman_starburst} was
tested. Loeb and Waxman assumed that protons in starburst galaxies with
energy less than 3 PeV convert almost all of their energy into pions. Their
work predicts a range that should encompass the true neutrino spectrum, but
the model tested here uses the most probable spectrum from the paper,
\mbox{$\Phi \propto$ E$^{-2.15}$}. This analysis assumed the flux was valid for energies
ranging from $10^{3}$ to \mbox{$10^{7}$ GeV}. The upper limit on this
spectral shape and normalization is 21.1$ \cdot \Phi_\mathrm{starburst}$.

\begin{figure*}
\centering
\includegraphics*[width=0.99\textwidth]{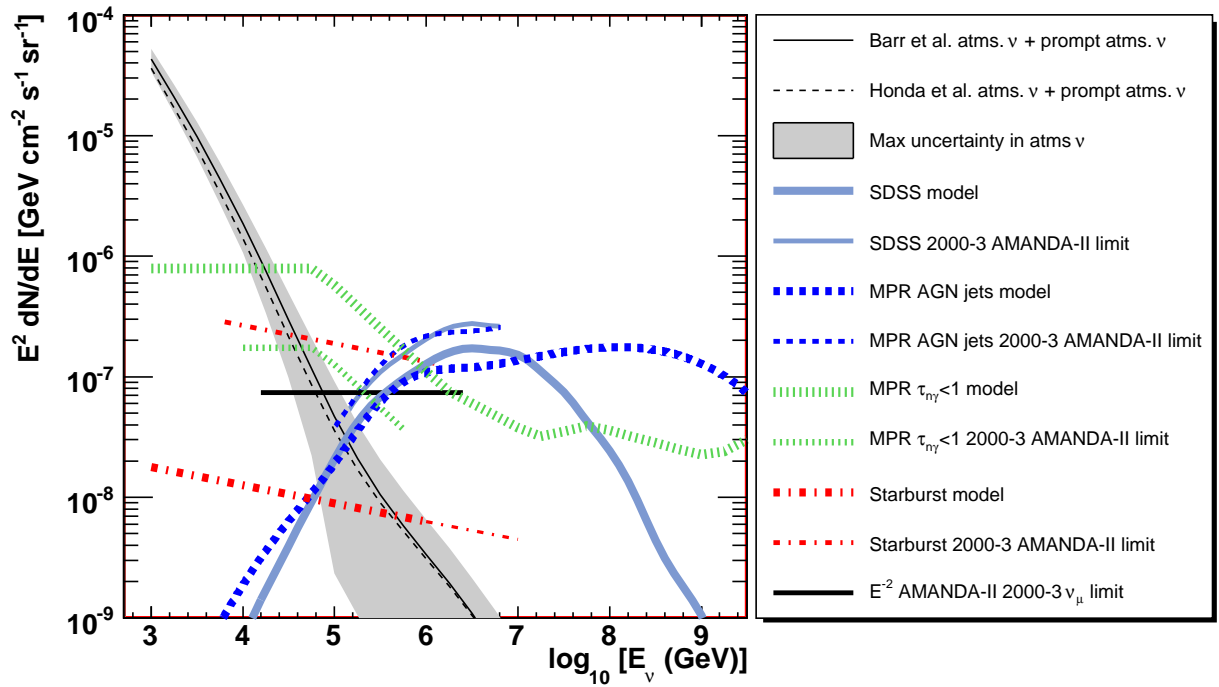}
\caption{\label{astroflux}Astrophysical neutrino models and upper
limits established with this analysis. The Barr \textit{et al.}  and Honda
\textit{et al.} atmospheric neutrino models are shown as thin lines with
maximum uncertainties assumed by this analysis represented by the
band. Other models that were tested included the SDSS AGN core model
\cite{sdss, sdss_revision}, the MPR upper bounds for AGN jets and optically
thin sources \cite{mpr}, and a starburst galaxy model
\cite{loeb_waxman_starburst}.}
\end{figure*}

These astrophysical neutrino models and their observed upper limits based
on this data set are shown in Figure \ref{astroflux}. Neutrino oscillations
are taken into account for all models where this factor was not already applied.

\subsubsection{\label{section_prompt}Prompt Neutrinos}

Since prompt neutrinos have a harder (less steep) spectrum than the
conventional atmospheric neutrinos, it is possible to search for a prompt
neutrino flux by separating the two event classes in energy. The final
$N_\mathrm{ch}$ requirement was reoptimized yielding $N_\mathrm{ch} \geq 71$ and the normalization factor
was determined based on the interval ($50 \leq N_\mathrm{ch} < 71$).

In the astrophysical neutrino searches described thus far, the range of
atmospheric neutrinos predicted in the final sample included an uncertainty
due to the unknown prompt neutrino flux. For the search for prompt
neutrinos, this uncertainty in the total atmospheric neutrino flux was
changed so that only conventional atmospheric neutrino uncertainties were
included. Since the atmospheric neutrino simulation was still normalized to
the low energy data, the overall effect in the atmospheric background
prediction for the final sample was small.

Martin \textit{et al.} predict prompt lepton fluxes based on the GBW model
for deep inelastic scattering. This model includes gluon saturation
effects \cite{martin_gbw} which lower the predicted charm production cross
sections. The predicted flux is lower than the sensitivity of this data
set. The upper limit on this model is $60.3 \cdot \Phi_\mathrm{Martin
GBW}$.

The Naumov RQPM \cite{naumov_rqpm_a,naumov_rqpm_b} model of prompt
atmospheric neutrinos incorporates data from primary cosmic ray and
hadronic interaction experiments. This non-perturbative model includes
intrinsic charm \cite{prompt_lepton_cookbook}. The upper limit on this
model is 5.2$ \cdot \Phi_\mathrm{Naumov RQPM}$.

Prompt neutrinos based on the models of Zas, Halzen and Vazquez were also
simulated \cite{zhv_charm}. A parameterization was established to describe
the energy dependence of the charm cross section. For the Charm C model,
the charm cross section was fitted to experimental data. In the Charm D
model, the cross section was parameterized by Volkova \cite{volkova}. Due
to the upward fluctuation in the number of events in the $N_\mathrm{ch}
\geq 71$ region, the upper limit for Charm C is 1.5$ \cdot
\Phi_\mathrm{Charm C}$. The upper limit on the Charm D model is 0.95$ \cdot
\Phi_\mathrm{Charm D}$. The MRF is less than 1.0, hence the Charm D model is
disfavored at the 90\% confidence level.

\begin{figure*}
\centering
\includegraphics*[width=0.99\textwidth]{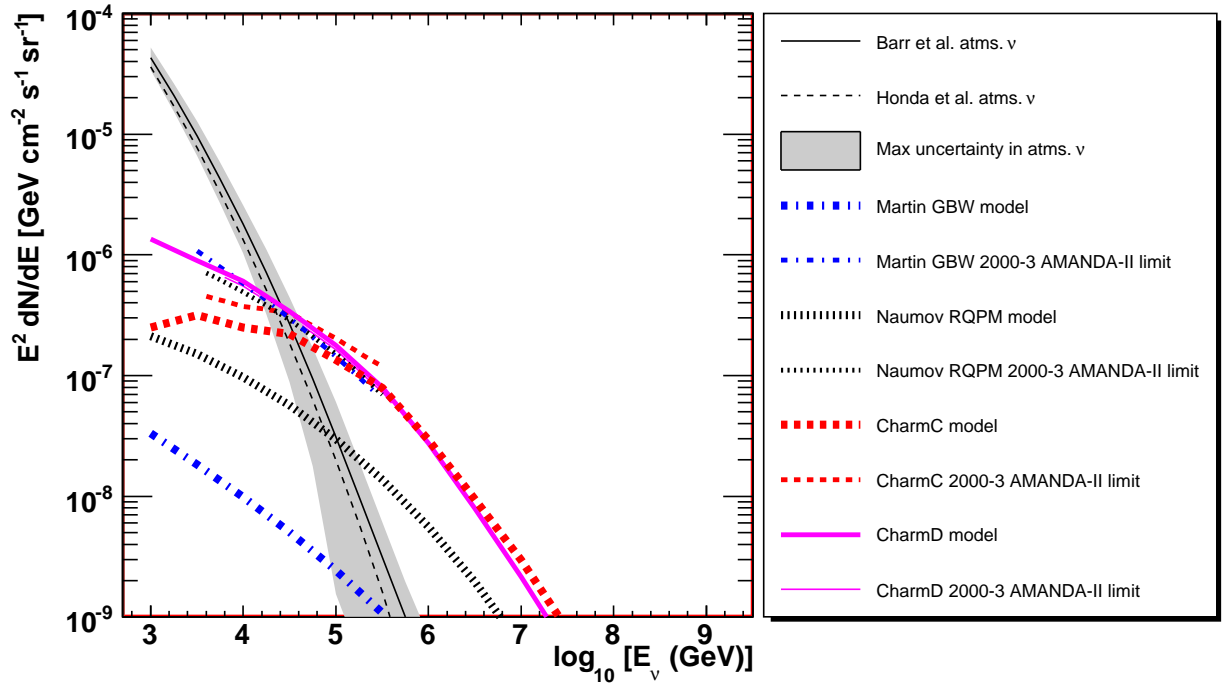}
\caption{\label{promptflux}Prompt neutrino models and upper limits based on this
analysis. The Barr \textit{et al.} and Honda
\textit{et al.} atmospheric neutrino predictions are shown
for reference. Two charm models \cite{zhv_charm} were tested, along with
the Naumov RQPM \cite{naumov_rqpm_a, naumov_rqpm_b} and Martin GBW
\cite{martin_gbw} models.}
\end{figure*}

The prompt neutrino models are shown in Figure \ref{promptflux}, along with
the upper limits based on these data.

\renewcommand{\arraystretch}{0.55}
\begin{table*}
\footnotesize
\begin{tabular} {|r|r|r|r|}
\multicolumn{4}{c}{\textbf{Astrophysical $\nu$}}\\
\hline
 & $\Phi \propto E^{-2}$ & SDSS \cite{sdss,sdss_revision} & MPR AGN jets
\cite{mpr} \\
\hline
$N_\mathrm{ch}$ & 100 & 139 & 139 \\
\hline
$n_\mathrm{b}$ & 7.0 & 1.55 & 1.55 \\
\hline
$n_\mathrm{s}$ & 66.7 & 1.74 & 1.42 \\
\hline
$\mu_\mathrm{median}(n_\mathrm{b})$ & 6.36 & 2.86 & 2.86 \\
\hline
sensitivity & & & \\
$\mu_\mathrm{median}(n_\mathrm{b})/n_\mathrm{s} \times \Phi$ & 0.095
$\times \Phi_\mathrm{E^{-2}}$ & 1.6 $\times \Phi_\mathrm{SDSS}$ & 2.0 $\times \Phi_\mathrm{MPR AGN}$ \\
\hline
$n_\mathrm{obs}$ & 6 & 1 & 1 \\
\hline
$\mu_{90\% \mathrm{ C.I.}}$ & (0,4.95) & (0,2.86) & (0,2.86) \\
\hline
upper limit & & & \\
$\mu/n_\mathrm{s} \times \Phi$ & 0.074 $\times \Phi_\mathrm{E^{-2}}$ & 1.6
$\times \Phi_\mathrm{SDSS}$ & 2.0 $\times \Phi_\mathrm{MPR AGN}$ \\
\hline
$(log_{10}E_{min},log_{10}E_{max}) $ & (4.2,6.4) & (5.1,6.8) & (5.0,6.9) \\
\hline

\end{tabular}

\begin{tabular} {|r|r|r|}
\multicolumn{3}{c}{}\\
\hline
 & MPR $\tau_{n\gamma} < 1$ \cite{mpr} & Starburst \cite{loeb_waxman_starburst} \\
\hline
$N_\mathrm{ch}$ & 86 & 71 \\
\hline
$n_\mathrm{b}$ & 12.9 & 29.1 \\
\hline
$n_\mathrm{s}$ & 42.7 & 1.05 \\
\hline
$\mu_\mathrm{median}(n_\mathrm{b})$ & 8.48 & 8.24 \\
\hline
sensitivity & & \\
$\mu_\mathrm{median}(n_\mathrm{b})/n_\mathrm{s} \times \Phi$ & 0.2 $\times
\Phi_\mathrm{MPR \tau<1}$ & 7.8 $\times \Phi_\mathrm{Starburst}$\\
\hline
$n_\mathrm{obs}$ & 14 & 37 \\
\hline
$\mu_{90\% \mathrm{ C.I.}}$ & (0,9.49) & (0,22.13) \\
\hline
upper limit & &\\
$\mu/n_\mathrm{s} \times \Phi$ & 0.22 $\times \Phi_\mathrm{MPR \tau<1}$ &
21.1 $\times \Phi_\mathrm{Starburst}$ \\
\hline
$(log_{10}E_{min},log_{10}E_{max}) $ & (4.0,5.8) & (3.8,6.1) \\
\hline
\end{tabular}

\begin{tabular} {|r|r|r|r|r|}
\multicolumn{5}{c}{\textbf{Prompt $\nu$}}\\
\hline
& Martin & Naumov & &  \\
 & GBW \cite{martin_gbw} & RQPM \cite{naumov_rqpm_a,naumov_rqpm_b} & CharmC
\cite{zhv_charm} & CharmD \cite{zhv_charm} \\
\hline
$N_\mathrm{ch}$ & 71 & 71 & 71 & 71 \\
\hline
$n_\mathrm{b}$ & 27.4 & 27.4 & 27.4 & 27.4 \\
\hline
$n_\mathrm{s}$ & 0.41 & 4.74 & 16.05 & 26.15 \\
\hline
$\mu_\mathrm{median}(n_\mathrm{b})$ & 8.75 & 8.75 & 8.75 & 8.75 \\
\hline
sensitivity & & & & \\
$\mu_\mathrm{median}(n_\mathrm{b})/n_\mathrm{s} \times \Phi$ & 21.3 $\times
\Phi_\mathrm{MGBW}$ & 1.8 $\times \Phi_\mathrm{NRQPM}$ & 0.55 $\times
\Phi_\mathrm{CharmC}$ & 0.33 $\times \Phi_\mathrm{CharmD}$\\
\hline
$n_\mathrm{obs}$ & 37 & 37 & 37 & 37 \\
\hline
$\mu_{90\% \mathrm{ C.I.}}$ & (1.29,24.72) & (1.29,24.72) & (1.29,24.72) & (1.29,24.72) \\
\hline
upper limit & & & & \\
$\mu/n_\mathrm{s} \times \Phi$ & 60.3 $\times \Phi_\mathrm{MGBW}$ & 5.2
$\times \Phi_\mathrm{NRQPM}$ & 1.5 $\times \Phi_\mathrm{CharmC}$ & 0.95 $\times
\Phi_\mathrm{CharmD}$ \\
\hline
$(log_{10}E_{min},log_{10}E_{max}) $ & (3.5,5.5) & (3.6,5.6) & (3.8,5.7) & (3.6,5.6) \\
\hline

\end{tabular}
\caption{\label{othermodelstable}Several flux shapes were tested with this
data set. $N_\mathrm{ch}$ is the minimum number of OMs that had to be hit
for an event to appear in the final data set. The predicted number of
events for background, $n_\mathrm{b}$, and signal, $n_\mathrm{s}$, were
determined by the simulation. The median event upper limit is
$\mu_\mathrm{median}(n_\mathrm{b})$. The sensitivity is the model flux
multiplied by the median event upper limit and divided by the number of
signal predicted. The number of events observed in the four year data
sample is $n_\mathrm{obs}$. The upper limit is calculated from the maximum
value of the 90\% confidence interval for the event upper limit, $\mu$. The
upper limit is the test flux multiplied by $\mu/n_\mathrm{s}$. All values
quoted here incorporate systematic uncertainties.}
\end{table*}
\renewcommand{\arraystretch}{1}

\section{\label{section_conclusions}Conclusions}

The experimental data were consistent with the predicted range of
atmospheric neutrino background. Six high energy events were observed in
the final data set, while the average predicted background was 7.0
events. There is no indication of an astrophysical signal. At a 90\%
confidence level, the diffuse flux of extraterrestrial muon neutrinos with
an E$^{-2}$ spectrum is not larger than \mbox{7.4 $\times$ 10$^{-8}$ GeV
cm$^{-2}$ s$^{-1}$ sr$^{-1}$} for \mbox{16 TeV} -- \mbox{2.5 PeV}.

This analysis also provides upper limits on four astrophysical neutrino
models and four prompt neutrino models. For the hardest signal spectra, the
results are consistent with background. The softer spectra were tested with
lower $N_\mathrm{ch}$ requirements and despite the observation leading to a
two-sided 90\% confidence interval, the level of excess is not significant
enough to claim a detection.

Before requiring events to fulfill $N_\mathrm{ch} \geq 100$, the observed
events were compared to the atmospheric neutrino simulation with systematic
uncertainties included. The observed low energy data were used to normalize
the atmospheric neutrino simulation, hence narrowing the range of
atmospheric neutrinos predicted by the different models for the final high
energy sample. Systematic effects of the event selection procedure were
studied in the inverted analysis using atmospheric muons. A consistency was
established between the observed downgoing atmospheric muon flux and the
upgoing atmospheric neutrino flux using the inverted analysis.

This result is the best upper limit on the diffuse flux of muon neutrinos
to date. The upper limit is an order of magnitude lower than the previous
AMANDA result by performing a multi-year analysis \cite{diffuse97} and by
using a larger detector, AMANDA-II instead of AMANDA-B10. For a $\Phi
\propto$ E$^{-2}$ spectral shape, this analysis provides an upper limit
that is a factor of three better than the Baikal muon neutrino upper limit
(muon neutrino upper limit = all-flavor limit/3 assuming a 1:1:1 flavor
ratio).

This analysis set upper limits on four prompt atmospheric neutrino
predictions, while one of these models is disfavored at a 90\% confidence
level. Other spectral shapes were tested for astrophysical neutrinos. No
models were excluded, however constraints were placed on the existing
predictions. The shapes of the MPR upper bounds were tested in the energy
region where the detector response peaks. For the benchmark $\Phi \propto$
E$^{-2}$ spectral shape, the current limit is a factor of 4 above the
Waxman-Bahcall upper bound.

AMANDA-II has now been integrated into IceCube. The sensitivity of the
IceCube detector will continue to improve as the detector grows to its
final volume, 1 km$^{3}$. Based on estimations
with AMANDA software, the full IceCube detector will have a sensitivity
that is a factor of 10 better than this analysis after one year of
operation \cite{icecubesensitivity}.

\renewcommand{\arraystretch}{0.65}
\begin{table*}
\footnotesize
\begin{tabular} {|r|r|r|r|r|r|r|}
\hline
\multicolumn{7}{|c|} {\textbf{Upgoing  0 \textless $N_\mathrm{ch}$ \textless
100}} \\
\hline
 & L0 & L1 & L2 & L3 & L4 & L5 \\
\hline
signal & & 979 & 664 & 486 & 437 & 141 \\
\hline
coinc. $\mu$ & & 60800 & 5750 & 530 & 248 & 0 \\
\hline
misreconstructed & & & & & & \\
CORSIKA & & & & & & \\
atms. $\mu$ & & 1340000 & 94900 & 1760 & 208 & 0 \\
\hline
Barr \textit{et al.} & & & & & & \\
atms. $\nu$ & & 9090 & 6590 & 4470 & 3890 & 534 \\
\hline
Honda \textit{et al.} & & & & & & \\
atms. $\nu$ & & 7290 & 5300 & 3600 & 3130 & 420 \\
\hline
Martin GBW & & 8.2 & 8.2 & 6.4 & 5.7 & 1.2 \\
prompt atms. $\nu$ & & & & & & \\
\hline
Naumov RQPM & & 71.6 & 71.6 & 56.7 & 50.7 & 11.5\\
prompt atms. $\nu$ & & & & & & \\
\hline
Data & & 3956810 & 294947 & 10841 & 4088 & 459 \\
\hline
\multicolumn{7}{|c|} {\textbf{Downgoing  0 \textless $N_\mathrm{ch}$ \textless
100}} \\
\hline
CORSIKA & & & & & & \\
atms. $\mu$ ($\times 10^{7}$) & 386 & & 288 & 212 & 195 & 24 \\
\hline
data ($\times 10^{7}$) & 432 & & 323 & 255 & 229 & 30 \\
\hline
\multicolumn{7}{r}{L5 = level of final analysis}\\
\end{tabular}
\caption{\label{invertedtable_less100} The number of low energy events (\mbox{0
\textless $N_\mathrm{ch}$ \textless 100}) at a given quality level for the different
types of simulation and experimental data.}
\end{table*}
\renewcommand{\arraystretch}{1}

\renewcommand{\arraystretch}{0.65}
\begin{table*}
\footnotesize
\begin{tabular} {|r|r|r|r|r|r|r|}
\hline
\multicolumn{7}{|c|} {\textbf{Upgoing  $N_\mathrm{ch} \geq 100$}} \\
\hline
 & L0 & L1 & L2 & L3 & L4 & L5 \\
\hline
signal &  & 160 & 124 & 104 & 103 & 68.4\\
\hline
coinc. $\mu$ & & 54.2 & 4.3 & 2.8 & 2.8 & 0  \\
\hline
misreconstructed & & & & & & \\
CORSIKA & & & & & & \\
atms. $\mu$ & & 862 & 35.4 & 0 & 0 & 0 \\
\hline
Barr \textit{et al.} & & & & & & \\
atms. $\nu$ & & 36.0 & 27.6 & 19.3 & 18.9 & 9.1 \\
\hline
Honda \textit{et al.} & & & & & & \\
atms. $\nu$ & & 25.2 & 19.3 & 13.5 & 13.2 & 6.4 \\
\hline
Martin GBW && 0.42 & 0.42 & 0.36 & 0.36 & 0.19 \\
prompt atms. $\nu$ & & & & & & \\
\hline
Naumov RQPM && 4.8 & 4.8 & 4.2 & 4.2 & 2.2 \\
prompt atms. $\nu$ & & & & & & \\
\hline
Data & & 11456 & 1347 & 96 & 45 & 6 \\
\hline
\multicolumn{7}{|c|} {\textbf{Downgoing  $N_\mathrm{ch} \geq 100$}} \\
\hline
CORSIKA & & & & & & \\
atms. $\mu$ ($\times 10^{7}$) & 7.31 & & 6.53 & 6.05 & 6.01 & 5.09 \\
\hline
data ($\times 10^{7}$) & 9.75 & & 8.59 & 8.07 & 8.03 & 6.60 \\
\hline
\multicolumn{7}{r}{L5 = level of final analysis}\\
\end{tabular}
\caption{\label{invertedtable_more100} The number of
high energy events (\mbox{$N_\mathrm{ch} \geq $100}) at a given quality level for
the different types of simulation and experimental data.}
\end{table*}
\renewcommand{\arraystretch}{1}

\begin{acknowledgments}

We acknowledge the support from the following agencies:
National Science Foundation-Office of Polar Program,
National Science Foundation-Physics Division,
University of Wisconsin Alumni Research Foundation,
Department of Energy, and National Energy Research Scientific Computing Center
(supported by the Office of Energy Research of the Department of Energy),
the NSF-supported TeraGrid system at the San Diego Supercomputer Center (SDSC),
and the National Center for Supercomputing Applications (NCSA);
Swedish Research Council,
Swedish Polar Research Secretariat,
and Knut and Alice Wallenberg Foundation, Sweden;
German Ministry for Education and Research,
Deutsche Forschungsgemeinschaft (DFG), Germany;
Fund for Scientific Research (FNRS-FWO),
Flanders Institute to encourage scientific and technological research in industry (IWT),
Belgian Federal Office for Scientific, Technical and Cultural affairs (OSTC);
the Netherlands Organisation for Scientific Research (NWO);
M.~Ribordy acknowledges the support of the SNF (Switzerland);
A. Kappes and J.~D.~Zornoza acknowledge support by the EU Marie Curie OIF Program.

\end{acknowledgments}

\newpage


\section*{Appendix A: Event Selection Techniques}

\begin{figure*}
\centering
\includegraphics*[width=3.4in]{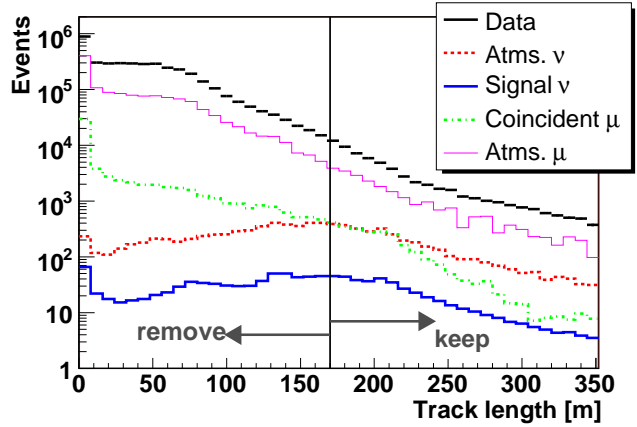}
\caption{\label{ldirbcut}The reconstructed track length within the detector
is shown. In order to identify muon neutrino tracks, events were required
to have long tracks of at least 170 meters. This removed a large fraction
of the atmospheric muon simulation, but had a smaller effect on the
atmospheric neutrino and signal simulations.}
\end{figure*}

Event selection techniques were applied to find the best reconstructed
upgoing tracks. The event requirements were tightened through a series of
values, becoming more restrictive at each of the five different levels. As
seen in Figure \ref{ldirbcut}, requiring a minimum value of the track
length, for instance, can be a powerful method of rejecting
misreconstructed downgoing backgrounds. The event selection requirements
for $L_\mathrm{dir}$, $N_\mathrm{dir}$, smoothness, median resolution and
likelihood ratio were established to remove many orders of magnitude more
misreconstructed background than upgoing atmospheric neutrinos or signal
neutrinos. Events which did not meet an optimized minimum or maximum value
of each parameter were removed.

The strength of these quality requirements was adjusted at each level. The
requirement is defined for each parameter in Table
\ref{leveldefinitions}. The plots in Figure
\ref{zenithplots} show the zenith angle distribution of all events
fulfilling the \mbox{zenith angle \textgreater $80^{\circ}$} and event
observable requirements at the chosen level. After the zenith angle
criteria was fulfilled at Level 1, the data mostly contains
misreconstructed atmospheric muons (top right, Figure
\ref{zenithplots}). As the quality parameters become more restrictive, the
data begins to follow the atmospheric neutrino simulation in the upgoing
direction and the atmospheric muon simulation in the downgoing
direction. At Level 5, the event quality requirements were strong enough to
have removed all of the misreconstructed downgoing atmospheric muon events
that were simulated. However, to be sure that the final data set only
included atmospheric and astrophysical neutrinos and no misreconstructed
downgoing events, an additional zenith angle requirement was imposed. All
events were kept if they were reconstructed between 100$^{\circ}$ and
180$^{\circ}$. The analysis continued with the data sample shown at
\mbox{Level 5}.

\renewcommand{\arraystretch}{0.65}
\begin{table*}
\scriptsize
\begin{tabular} {|c|c|c|c|c|c|c|}
\hline
 & L0 & L1 & L2 & L3 & L4 & L5* \\
\hline
\textbf{Zenith Angle} [$^o$] & & \textgreater80 & \textgreater80 &
\textgreater80 & \textgreater80 & \textgreater100 \\
\hline
\textbf{Number of} & & & & & & \\
\textbf{Direct Hits} & & & \textgreater5 & \textgreater8 & \textgreater8 &
\textgreater13 \\
\hline
\textbf{Track Length} [m]  & & & \textgreater100 & \textgreater130 & \textgreater130 & \textgreater170 \\
\hline
\textbf{$|$Smoothness$|$} & & & & \textless$0.30$ & \textless$0.30$ & \textless$0.25$ \\
\hline
\textbf{Median} & & & & & & \\
\textbf{Resolution} [$^o$] & & & & & \textless4.0 & \textless4.0 \\
\hline
\textbf{Likelihood} & & & & & & \\
\textbf{Ratio ($\Delta\mathnormal{L}$)} & & & & & & \\
\textbf{vs. Zenith} & & & & & & $\Delta\mathnormal{L}>-38.2~cos(zenith)+27.506$\\
\hline
\hline
\textbf{Number of} & & & & & & \\
\textbf{Remaining Events} &  $5.2 \times 10^{9}$ & $7.8 \times 10^{6}$ &
$1.2 \times 10^{6}$ & $3.5 \times 10^{5}$ & $1.8 \times 10^{5}$ & 465\\
\hline
\multicolumn{7}{r} {* = level of the final analysis}\\
\end{tabular}
\caption{\label{leveldefinitions} The table summarizes the event quality
requirements as a function of quality level. Events only remained in the
sample if they fulfilled all of the parameter requirements for a given
level. The removal of all horizontal events (zenith $<$ 100) contributed to
the large decrease in events from L4 to L5.}
\end{table*}
\renewcommand{\arraystretch}{1}

\pagebreak[4]

\section*{Appendix B: Neutrino Effective Area}

\renewcommand{\arraystretch}{0.65}
\begin{table*}[!h]
\scriptsize
\begin{tabular} {|c||c|c||c|c||}
\hline
\textbf{Energy} & \multicolumn{2}{|c||} {\textbf{-1 \textless cos(Zenith) \textless -.8}} & \multicolumn{2}{|c||}
{\textbf{-.8 \textless cos(Zenith) \textless -.6}} \\
 log$_{10}$  (E/GeV)  & $\nu_{\mu}$ & $\bar{\nu}_{\mu}$ & $\nu_{\mu}$ & $\bar{\nu}_{\mu}$ \\
 & [$10^{3} cm^{2}$] & [$10^{3} cm^{2}$] &  [$10^{3} cm^{2}$] &
[$10^{3} cm^{2}$] \\
\hline
\hline
3.6 & 0.487 & 0.166 & 0.279 & 0.0673 \\
\hline
3.8 & 1.04 & 1.1 & 0.652 & 0.646 \\
\hline
4 & 3.36 & 2.85 & 1.82 & 1.89 \\
\hline
4.2 & 8.74 & 7.54 & 4.97 & 5.56 \\
\hline
4.4 & 18.8 & 16.2 & 15.3 & 12.4 \\
\hline
4.6 & 29.3 & 30.4 & 34 & 26.9 \\
\hline
4.8 & 44.9 & 46.4 & 52.7 & 58.8 \\
\hline
5 & 59.6 & 65.5 & 92.6 & 88 \\
\hline
5.2 & 75.7 & 69.7 & 128 & 121 \\
\hline
5.4 & 72.6 & 84.4 & 153 & 163 \\
\hline
5.6 & 63.5 & 77.8 & 180 & 179 \\
\hline
5.8 & 63.3 & 66.9 & 183 & 188 \\
\hline
6 & 51.9 & 49.3 & 170 & 177 \\
\hline
6.2 & 36.6 & 39.1 & 145 & 151 \\
\hline
6.4 & 27.8 & 22.6 & 110 & 113 \\
\hline
6.6 & 9.97 & 14.7 & 72.3 & 77 \\
\hline
6.8 & 7.8 & 8.73 & 54.2 & 48.2 \\
\hline
7 & 3.39 & 3.08 & 29.6 & 29.5 \\
\hline
7.2 & 3.12 & 1.44 & 16.5 & 15.2 \\
\hline
7.4 & 0.939 & 0.718 & 7.97 & 9.64 \\
\hline
7.6 & 0.864 & 0.791 & 5.12 & 4.15 \\
\hline
7.8 & 0.492 & 0.521 & 2.59 & 2.08 \\
\hline
\end{tabular}

\begin{tabular} {|c||c|c||c|c||}
\multicolumn{5} {c} {} \\ 
\hline
\textbf{Energy} & \multicolumn{2}{|c||} {\textbf{-.6 \textless cos(Zenith) \textless -.4}} & \multicolumn{2}{|c||}
{\textbf{-.4 \textless cos(Zenith) \textless -.17}} \\
 log$_{10}$  (E/GeV)  & $\nu_{\mu}$ & $\bar{\nu}_{\mu}$ & $\nu_{\mu}$ & $\bar{\nu}_{\mu}$ \\
 & [$10^{3} cm^{2}$] & [$10^{3} cm^{2}$] &  [$10^{3} cm^{2}$] &
[$10^{3} cm^{2}$] \\
\hline
\hline
3.6 & 0.108 & 0.0562 & 0.0752 & 0.0451 \\
\hline
3.8 & 0.282 & 0.163 & 0.178 & 0.0818 \\
\hline
4 & 0.845 & 0.93 & 1.13 & 0.543 \\
\hline
4.2 & 3.73 & 3.39 & 1.98 & 1.66 \\
\hline
4.4 & 9.74 & 8.22 & 7.23 & 6.02 \\
\hline
4.6 & 21.1 & 19.9 & 17.9 & 18.2 \\
\hline
4.8 & 49.7 & 43.3 & 33.2 & 36.9 \\
\hline
5 & 86.2 & 77.5 & 74.2 & 68.3 \\
\hline
5.2 & 118 & 119 & 119 & 113 \\
\hline
5.4 & 179 & 165 & 163 & 167 \\
\hline
5.6 & 232 & 217 & 264 & 230 \\
\hline
5.8 & 243 & 232 & 306 & 310 \\
\hline
6 & 271 & 286 & 377 & 373 \\
\hline
6.2 & 269 & 258 & 418 & 389 \\
\hline
6.4 & 251 & 229 & 441 & 452 \\
\hline
6.6 & 212 & 197 & 437 & 391 \\
\hline
6.8 & 154 & 149 & 417 & 437 \\
\hline
7 & 105 & 114 & 413 & 380 \\
\hline
7.2 & 79.8 & 61.4 & 328 & 327 \\
\hline
7.4 & 46.3 & 32.9 & 285 & 274 \\
\hline
7.6 & 31.8 & 19.4 & 209 & 212 \\
\hline
7.8 & 17.7 & 10.3 & 142 & 146 \\
\hline
\end{tabular}

\caption{\label{effareatable}Effective area as a function of the energy and
zenith angle of the simulation.}
\end{table*}
\renewcommand{\arraystretch}{1}

\renewcommand{\arraystretch}{0.65}
\begin{table*}
\small
\begin{tabular} {|c||c|c||}
\multicolumn{3} {l} {\textbf{Effective Area in cm$^{2}$}} \\ 
\multicolumn{3} {l} {} \\ 
\hline
\textbf{Energy} & \multicolumn{2}{|c||} {\textbf{All angle}} \\
 log$_{10}$  (E/GeV)  & $\nu_{\mu}$ & $\bar{\nu}_{\mu}$ \\
 & [$10^{3} cm^{2}$] & [$10^{3} cm^{2}$] \\
\hline
\hline
3.6 & 0.164 & 0.0572 \\
\hline
3.8 & 0.381 & 0.343 \\
\hline
4 & 1.24 & 1.07 \\
\hline
4.2 & 3.33 & 3.15 \\
\hline
4.4 & 8.9 & 7.51 \\
\hline
4.6 & 17.9 & 16.7 \\
\hline
4.8 & 31.8 & 32.6 \\
\hline
5 & 55.6 & 52.7 \\
\hline
5.2 & 78.9 & 75.8 \\
\hline
5.4 & 102 & 103 \\
\hline
5.6 & 136 & 127 \\
\hline
5.8 & 144 & 145 \\
\hline
6 & 161 & 162 \\
\hline
6.2 & 164 & 155 \\
\hline
6.4 & 157 & 155 \\
\hline
6.6 & 139 & 130 \\
\hline
6.8 & 121 & 126 \\
\hline
7 & 112 & 102 \\
\hline
7.2 & 86 & 83.4 \\
\hline
7.4 & 67.8 & 63.6 \\
\hline
7.6 & 51.2 & 49 \\
\hline
7.8 & 35.5 & 34.6 \\
\hline
\end{tabular}
\caption{\label{effareatable_angleaveraged} The angle-averaged neutrino
effective area as a function of energy.}
\end{table*}
\renewcommand{\arraystretch}{1}

\pagebreak[4]

\section*{Appendix C: Atmospheric Neutrino Flux}

For this analysis, the atmospheric neutrino flux models by Barr \textit{et
al.} and Honda \textit{et al.} were both considered equally likely options
for the background atmospheric neutrino simulation. These two models are
among many that use slightly different parameterizations of the all-nucleon
cosmic ray flux to derive the atmospheric neutrino flux.

For this analysis, the Barr \textit{et al.} flux below 10 GeV was taken
from \cite{bartol2004}. From 10 GeV to 10 TeV, the flux tables from
\cite{barronlinetables}, based on the primary spectrum of
\cite{icrc2001gaisser}, were used. Above 10 TeV, the weight was
derived by performing a 2-dimensional fit with a fifth degree polynomial to
the log$_{10}$E vs. cos(zenith) tables of the atmospheric neutrino flux
values from lower energies just mentioned. The TARGET version 2.1
\cite{target} hadronic interaction model was used \cite{bartol2004}.

In an attempt to better fit the AMS \cite{ams} and BESS
\cite{bess,bess-tev} data, Honda \emph{et al.} changed the power law fit to
the proton cosmic ray spectrum from -2.74 to -2.71 above 100
GeV \cite{honda2004}. Other parameters in the cosmic ray fit remained
similar to the Barr \emph{et al.} flux mentioned above \cite{gaisser2005},
although the DPMJET-III \cite{dpmjet} interaction model was used. The
atmospheric neutrino weights from \cite{honda2004} were used up to 10
TeV. Above that energy, a 2-dimensional fit of the lower energy values was
again used as described above. The result was a lower atmospheric neutrino
flux prediction than the Barr \emph{et al.} flux.

\begin{figure*}
\centering
\includegraphics*[width=3.4in]{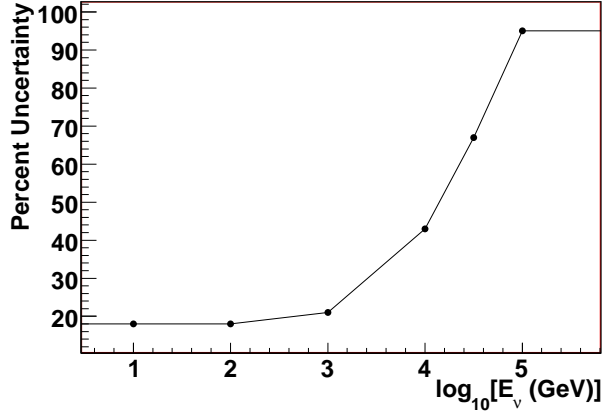}
\caption{\label{energyuncertainty}The estimated uncertainty in the
atmospheric neutrino flux as a function of energy. Due to the large
uncertainty in the prompt neutrino flux at greater than 10$^{4}$ GeV, the
total uncertainty rises sharply.}
\end{figure*}

As described in Section \ref{subsection_theoretical_unc}, uncertainties in
hadronic interactions and the cosmic ray and prompt neutrino fluxes at high
energy led to large total uncertainties in the atmospheric neutrino
flux. The estimated uncertainties are shown in Figure
\ref{energyuncertainty}.

The atmospheric neutrino simulation was renormalized based on the
experimental low energy data. The number of low energy conventional
atmospheric neutrinos (second column) is added to the 4.0 prompt neutrinos
predicted with the central prompt neutrino model. The total atmospheric
background prediction before renormalization is shown in the third column of
Table \ref{atmsnueventtable}. Instead of renormalizing the simulation based
on all events with \mbox{$N_\mathrm{ch} <$ 100}, the renormalization was
only based on the region \mbox{50 $<$ $N_\mathrm{ch}$ $<$ 100}. Because of
the difficulty of simulating events near the threshold of the detector, the
atmospheric neutrino simulation did not faithfully reproduce the shape of
the $N_\mathrm{ch}$ distribution for the data at $N_\mathrm{ch}$ below
50. Atmospheric neutrino models were scaled to match the 146 events seen in
the experimental data for \mbox{50 $<$ $N_\mathrm{ch}$ $<$ 100}. The total number
of high energy events predicted to survive the final energy requirement is
shown before renormalization in the sixth column and after renormalization
in the last column.

\renewcommand{\arraystretch}{0.65}
\begin{table*}
\scriptsize
\begin{tabular} {|r|c|c|c|c|c|c|}
\hline
\textbf{Atms. $\nu $ Model} & \textbf{Conv.} & \textbf{Conv. $\nu$ +} &
\textbf{Scale Factor} & \textbf{Conv.} & \textbf{Conv. $\nu$ +} &
\textbf{Background}\\

 & \textbf{Atms. $\nu$} & \textbf{prompt $\nu$} & \textbf{to 146} &
\textbf{Atms. $\nu$} & \textbf{prompt $\nu$} & \textbf{Predicted in}\\

 & 50 \textless $N_\mathrm{ch}$ \textless 100 &  50 \textless $N_\mathrm{ch}$ \textless 100  & \textbf{Low Energy} &
$N_\mathrm{ch} \geq 100$ & $N_\mathrm{ch} \geq 100$ & \textbf{ $N_\mathrm{ch} \geq 100$}\\

 & &  & \textbf{Data Events} & & & \textbf{Sample} \\
 & & & & & & \textbf{after Scaling} \\
\hline  
Barr \emph{et al.} Max & 249 & 253 & 0.58 & 13.3 & 14.5 & 8.3\\
\hline
Barr  \emph{et al.} & 194 & 198 & 0.74 & 9.1 & 10.3 & 7.6\\
\hline
Barr \emph{et al.}  Min & 138 & 142 & 1.03 & 4.9 & 6.1 & 6.3\\
\hline
Honda \emph{et al.} Max & 191 & 195 & 0.75 & 9.3 & 10.5 & 7.9\\
\hline
Honda \emph{et al.} & 149 & 153 & 0.96 & 6.4 & 7.6 & 7.3\\
\hline
Honda \emph{et al.} Min & 107 & 111 & 1.32 & 3.4 & 4.6 & 6.1\\
\hline
\end{tabular}
\caption{\label{atmsnueventtable}Number of atmospheric neutrino events predicted by the Monte
Carlo. Uncertainty in the high energy cosmic ray flux was incorporated into
the maximum and minimum predictions.}
\end{table*}
\renewcommand{\arraystretch}{1}

\pagebreak[4]

The following pages are an erratum to the original document.

\pagebreak[4]

\section*{Erratum: Multi-year search for a diffuse flux of muon neutrinos with
AMANDA-II}

A search for TeV -- PeV muon neutrinos with \mbox{AMANDA-II} data collected
between 2000 and 2003 established an upper limit of
\mbox{E$^{2}\Phi_\mathrm{90\% C.L.} < 7.4 \times 10^{-8}$ GeV cm$^{-2}$
s$^{-1}$ sr$^{-1}$} on the diffuse flux of extraterrestrial
muon neutrinos with a {$\Phi \propto$ E$^{-2}$} spectrum between 16 TeV and
2.5 PeV. The upper limit calculation correctly included event simulations
and remains as stated. However, the calculation of the detector's efficiency, which is 
based only on simulations, was incorrectly tabulated in an appendix and
shown in a figure. The values were approximately a factor of ten too high, although the exact error varies in each bin. 
The correction has been applied in the following tables and figure. The effective area is the equivalent area over which the
detector would be 100\% efficient for detecting neutrinos. The typical uncertainty on the effective area from simulation statistics is lowest between 10$^{5}$ GeV and 10$^{6}$ GeV (2\%). The uncertainty increases to 6\% at 10$^{4}$ GeV and 5\% around 10$^{7}$ GeV. In the remainder of this document, the number of optical modules (OMs) triggered during an event is referred to as 
$N_\mathrm{ch}$ and cos($\theta_\mathrm{t}$) refers to the cosine of the simulated (true) zenith angle of an event. The term angle-averaged indicates that results are averaged over $\theta_\mathrm{t}$ between 100$^{\circ}$ and 180$^{\circ}$. All other
results reported in the paper, including the upper limit, remain unchanged.

\renewcommand{\arraystretch}{0.65}
\begin{table*}[!h]
\scriptsize
\begin{tabular} {|c||c|c||c|c||}
\hline
\textbf{Energy} & \multicolumn{2}{|c||} {\textbf{-1.0 \textless cos($\theta_\mathrm{t}$) \textless -0.8}} & \multicolumn{2}{|c||}
{\textbf{-0.8 \textless cos($\theta_\mathrm{t}$) \textless -0.6}} \\
 log$_{10}$  (E/GeV)  & $\nu_{\mu}$ & $\bar{\nu}_{\mu}$ & $\nu_{\mu}$ & $\bar{\nu}_{\mu}$ \\
 & [$10^{3} \mathrm{cm}^{2}$] & [$10^{3} \mathrm{cm}^{2}$] &  [$10^{3} \mathrm{cm}^{2}$] &
[$10^{3} \mathrm{cm}^{2}$] \\
\hline
\hline
3.6 & 0.046 & 0.017 & 0.024 & 0.0084 \\ 
\hline 
3.8 & 0.094 & 0.1 & 0.052 & 0.049 \\ 
\hline 
4 & 0.32 & 0.29 & 0.19 & 0.18 \\ 
\hline 
4.2 & 0.81 & 0.74 & 0.48 & 0.52 \\ 
\hline 
4.4 & 1.7 & 1.5 & 1.4 & 1.1 \\ 
\hline 
4.6 & 2.6 & 2.7 & 2.9 & 2.5 \\ 
\hline 
4.8 & 4 & 4 & 4.8 & 5.2 \\ 
\hline 
5 & 5.3 & 5.7 & 8.2 & 7.6 \\ 
\hline 
5.2 & 6.5 & 6.2 & 11 & 11 \\ 
\hline 
5.4 & 6.4 & 7.4 & 14 & 14 \\ 
\hline 
5.6 & 5.6 & 6.4 & 16 & 16 \\ 
\hline 
5.8 & 5.2 & 6 & 16 & 16 \\ 
\hline 
6 & 4.3 & 4.3 & 15 & 15 \\ 
\hline 
6.2 & 3.3 & 3.3 & 13 & 13 \\ 
\hline 
6.4 & 2.4 & 2 & 9.4 & 9.7 \\ 
\hline 
6.6 & 0.91 & 1.2 & 6.5 & 6.6 \\ 
\hline 
6.8 & 0.71 & 0.66 & 4.3 & 4.2 \\ 
\hline 
7 & 0.37 & 0.28 & 2.6 & 2.7 \\ 
\hline 
7.2 & 0.26 & 0.15 & 1.5 & 1.5 \\ 
\hline 
7.4 & 0.078 & 0.07 & 0.83 & 0.87 \\ 
\hline 
7.6 & 0.074 & 0.047 & 0.45 & 0.49 \\ 
\hline 
7.8 & 0.02 & 0.055 & 0.26 & 0.19 \\ 
\hline
\hline
\end{tabular}

\begin{tabular} {|c||c|c||c|c||}
\multicolumn{5} {c} {} \\ 
\hline
\textbf{Energy} & \multicolumn{2}{|c||} {\textbf{-0.6 \textless cos($\theta_\mathrm{t}$) \textless -0.4}} & \multicolumn{2}{|c||}
{\textbf{-0.4 \textless cos($\theta_\mathrm{t}$) \textless -0.17}} \\
 log$_{10}$  (E/GeV)  & $\nu_{\mu}$ & $\bar{\nu}_{\mu}$ & $\nu_{\mu}$ & $\bar{\nu}_{\mu}$ \\
 & [$10^{3} \mathrm{cm}^{2}$] & [$10^{3} \mathrm{cm}^{2}$] &  [$10^{3} \mathrm{cm}^{2}$] &
[$10^{3} \mathrm{cm}^{2}$] \\
\hline
\hline
3.6 & 0.0087 & 0.0043 & 0.0055 & 0.0032 \\ 
\hline 
3.8 & 0.035 & 0.018 & 0.015 & 0.01 \\ 
\hline 
4 & 0.081 & 0.087 & 0.11 & 0.037 \\ 
\hline 
4.2 & 0.35 & 0.31 & 0.16 & 0.14 \\ 
\hline 
4.4 & 0.9 & 0.8 & 0.69 & 0.59 \\ 
\hline 
4.6 & 1.9 & 1.9 & 1.5 & 1.6 \\ 
\hline 
4.8 & 4.4 & 4.1 & 3 & 3.2 \\ 
\hline 
5 & 7.5 & 7.1 & 6.8 & 5.8 \\ 
\hline 
5.2 & 11 & 11 & 11 & 10 \\ 
\hline 
5.4 & 16 & 14 & 15 & 14 \\ 
\hline 
5.6 & 20 & 19 & 23 & 20 \\ 
\hline 
5.8 & 22 & 20 & 26 & 27 \\ 
\hline 
6 & 23 & 24 & 32 & 32 \\ 
\hline 
6.2 & 24 & 23 & 37 & 33 \\ 
\hline 
6.4 & 22 & 20 & 38 & 38 \\ 
\hline 
6.6 & 18 & 17 & 37 & 34 \\ 
\hline 
6.8 & 13 & 13 & 36 & 37 \\ 
\hline 
7 & 9.4 & 9.8 & 34 & 31 \\ 
\hline 
7.2 & 6.9 & 5.9 & 27 & 29 \\ 
\hline 
7.4 & 4 & 3.4 & 23 & 23 \\ 
\hline 
7.6 & 2.7 & 1.7 & 16 & 18 \\ 
\hline 
7.8 & 1.1 & 1.1 & 12 & 13 \\ 
\hline 
\hline
\end{tabular}

\caption{\label{effareatable}Effective area as a function of the energy and
zenith angle of the simulation for events in the final sample satisfying \mbox{$N_\mathrm{ch} \geq 100$}.}
\end{table*}
\renewcommand{\arraystretch}{1}

\renewcommand{\arraystretch}{0.65}
\begin{table*}
\small
\begin{tabular} {|c||c|c||}
\hline
\textbf{Energy} & \multicolumn{2}{|c||} {\textbf{Angle-averaged}} \\
 log$_{10}$  (E/GeV)  & $\nu_{\mu}$ & $\bar{\nu}_{\mu}$ \\
 & [$10^{3} \mathrm{cm}^{2}$] & [$10^{3} \mathrm{cm}^{2}$] \\
\hline
\hline
3.6 & 0.02 & 0.0081 \\ 
\hline 
3.8 & 0.048 & 0.044 \\ 
\hline 
4 & 0.17 & 0.14 \\ 
\hline 
4.2 & 0.44 & 0.42 \\ 
\hline 
4.4 & 1.1 & 0.99 \\ 
\hline 
4.6 & 2.2 & 2.1 \\ 
\hline 
4.8 & 4 & 4.1 \\ 
\hline 
5 & 6.9 & 6.5 \\ 
\hline 
5.2 & 9.7 & 9.5 \\ 
\hline 
5.4 & 13 & 13 \\ 
\hline 
5.6 & 16 & 15 \\ 
\hline 
5.8 & 18 & 18 \\ 
\hline 
6 & 19 & 19 \\ 
\hline 
6.2 & 20 & 19 \\ 
\hline 
6.4 & 19 & 18 \\ 
\hline 
6.6 & 16 & 16 \\ 
\hline 
6.8 & 14 & 15 \\ 
\hline 
7 & 12 & 12 \\ 
\hline 
7.2 & 9.5 & 9.9 \\ 
\hline 
7.4 & 7.5 & 7.3 \\ 
\hline 
7.6 & 5.3 & 5.6 \\
\hline 
7.8 & 3.5 & 3.9 \\ 
\hline 
\end{tabular}
\caption{\label{effareatable_angleaveraged} The angle-averaged neutrino
effective area as a function of energy for events in the final sample satisfying \mbox{$N_\mathrm{ch} \geq 100$}.}
\end{table*}
\renewcommand{\arraystretch}{1}

\pagebreak[4]

\begin{figure*}
\centering
\includegraphics*[width=3.4in]{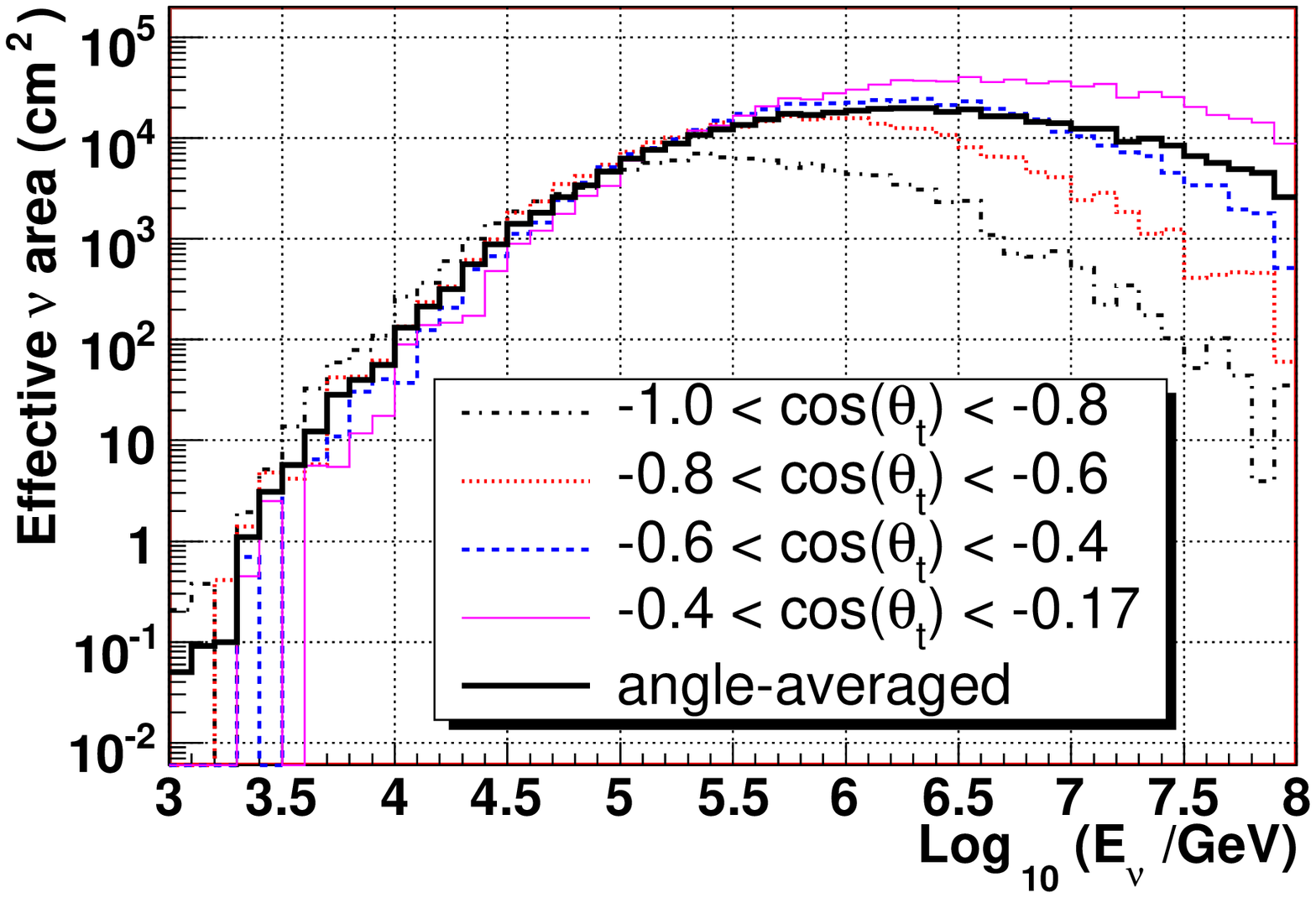}
\caption{\label{effareaplot}Effective area for $\nu_{\mu}$ as a function
of the true simulated energy at the Earth's surface in intervals of cosine
of the true zenith angle, $\theta_\mathrm{t}$. The angle-averaged effective area is represented by the solid black line. This calculation was based on the final event sample for events satisfying \mbox{$N_\mathrm{ch} \geq 100$}.}
\end{figure*}


\begin{thebibliography}{99}

\bibitem{wb_bound}
    E. Waxman and J. Bahcall, Phys. Rev. D \textbf{59}, 023002 (1998).

\bibitem{wb_robust}
    J. Bahcall and E. Waxman, Phys. Rev. D \textbf{64}, 023002 (2001).

\bibitem{wb_withredshift}
    E. Waxman, Nucl. Phys. Proc. Suppl. \textbf{118}, 353 (2003).

\bibitem{nellen}
    L. Nellen, K. Mannheim, and P.L. Biermann, Phys. Rev. D \textbf{47}, 5270 (1993).

\bibitem{becker}
    J. Becker, P. Biermann, and W. Rhode, Astropart. Phys. \textbf{23}, 355 (2005).

\bibitem{mpr}
    K. Mannheim, R.J. Protheroe, and J.P. Rachen, Phys. Rev. D \textbf{63},
023003 (2000).

\bibitem{sdss}
    F.W. Stecker, C. Done, M.H. Salamon, and P. Sommers,
Phys. Rev. Lett. \textbf{66}, 2697 (1991); \textbf{69}, 2738(E) (1992).

\bibitem{sdss_revision}
    F.W. Stecker, Phys. Rev. D \textbf{72}, 107301 (2005).

\bibitem{loeb_waxman_starburst}
    A. Loeb and E. Waxman, J. Cosmol. Astropart. Phys. JCAP05 003 (2006).

\bibitem{diffuse97}
    J. Ahrens \emph{et al.}, Phys. Rev. Lett. \textbf{90}, 251101 (2003).

\bibitem{cascades2000}
    M. Ackermann \emph{et al.}, Astropart. Phys. \textbf{22}, 127 (2004).

\bibitem{frejus}
    W. Rhode \emph{et al.} (Fr\'ejus Collaboration), Astropart. Phys. \textbf{4}, 217 (1996).

\bibitem{macro}
    M. Ambrosio \emph{et al.} (MACRO Collaboration), Astropart. Phys. \textbf{19}, 1 (2003).

\bibitem{baikal}
    V. Aynutdinov \emph{et al.}, Astropart. Phys. \textbf{25}, 140 (2006).

\bibitem{athar}
   H. Athar, M. Jezabek, and O. Yasuda, Phys. Rev. D \textbf{62}, 103007 (2000).

\bibitem{pointsource5yr}
    A. Achterberg \emph{et al.}, astro-ph 0611063 (2006).

\bibitem{nim2004}
    J. Ahrens \emph{et al.} (AMANDA Collaboration), Nucl. Instr. Meth. A \textbf{524}, 169 (2004).

\bibitem{corsika}
    D. Heck, J. Knapp, J.N. Capdevielle, G. Schatz, and T. Thouw,
Tech. Rep. FZKA 6019, Forschungszentrum Karlsruhe (1998).

\bibitem{nusim}
    G.C. Hill,  Astropart. Phys. \textbf{6}, 215 (1997). 

\bibitem{honda2004}
    M. Honda, T. Kajita, K. Kasahara, and S. Midorikawa, Phys. Rev. D \textbf{70},
043008 (2004).

\bibitem{bartol2004}
    G.D. Barr, T.K. Gaisser, P. Lipari, S. Robbins, and T. Stanev,
Phys. Rev. D \textbf{70}, 023006 (2004).

\bibitem{barronlinetables}
    G. Barr, T.K. Gaisser, P. Lipari, S. Robbins, and T. Stanev, http://www-pnp.physics.ox.ac.uk/$\sim$barr/fluxfiles/.

\bibitem{icrc2001gaisser}
    T.K. Gaisser, M. Honda, P. Lipari, and T. Stanev, in \emph{Proceedings of the 27th International Cosmic Ray
Conference}, Hamburg, Germany, \textbf{5}, 1643 (2001).

\bibitem{lipari}
    P. Lipari, Astropart. Phys. \textbf{1}, 195 (1993).

\bibitem{till_medres}
    T. Neunhoffer, Astropart. Phys. \textbf{25}, 220 (2006).

\bibitem{mrp}
    G.C. Hill and K. Rawlins, Astropart. Phys. \textbf{19}, 393 (2003).

\bibitem{feldcous}
    G.J. Feldman and R.D. Cousins, Phys. Rev. D \textbf{57}, 3873 (1998).

\bibitem{lisa_icrc2005}
    L. Gerhardt for the IceCube Collaboration, in \emph{Proceedings of the 29th International Cosmic Ray
Conference}, Pune, India, \textbf{5}, 111 (2005), astro-ph 0509330.

\bibitem{icepaper}
    M. Ackermann \emph{et al.}, J. Geophys. Res. \textbf{111}, D13203 (2006).

\bibitem{martin_gbw}
    A.D. Martin, M.G. Ryskin, and A.M. Stasto, Acta Phys. Polon. \textbf{B34}, 3273
(2003).

\bibitem{naumov_rqpm_a}
    G. Fiorentini, A. Naumov, and F.L. Villante, Phys. Lett. B \textbf{510}, 173 (2001).

\bibitem{naumov_rqpm_b}
    E.V. Bugaev \emph{et al.}, Il Nuovo Cimento \textbf{12C}, No. 1, 41 (1989).

\bibitem{prompt_lepton_cookbook}
    C.G.S. Costa, Astropart. Phys. \textbf{16}, 193 (2001).

\bibitem{zhv_charm}
    E. Zas, F. Halzen, and R.A. V\'{a}zquez, Astropart. Phys. \textbf{1}, 297 (1993).


\bibitem{gaisser_honda_review}
    T.K. Gaisser and M. Honda, Annu. Rev. Nucl. Part. Sci. \textbf{52},
153 (2002).

\bibitem{gaisser2005}
    T.K. Gaisser, in \emph{Proceedings of
Nobel Symposium 129 ``Neutrino Physics''}, astro-ph 0502380 (2005).

\bibitem{dcorsikaneutrinos}
    D. Chirkin, hep-ph 0407078 (2004).

\bibitem{cousinshighland}
    R.D. Cousins and V.L. Highland, Nucl. Instrum. Methods Phys. Res. A
\textbf{320}, 331 (1992).

\bibitem{conrad}
    J. Conrad, O. Botner, A. Hallgren, and C.~P\'erez~de~los~Heros, Phys. Rev. D \textbf{67}, 012002 (2003).

\bibitem{hillci}
    G.C. Hill, Phys. Rev. D \textbf{67}, 118101 (2003).

\bibitem{volkova}
    L.V. Volkova \emph{et al.}, Il Nuovo Cimento \textbf{C10}, 465 (1987).

\bibitem{kirsten_unfolding}
    K. M\"unich for the IceCube Collaboration, in \textit{Proceedings of the
29th International Cosmic Ray Conference}, Pune, India, \textbf{5}, 17
(2005), astro-ph 0509330.

\bibitem{uhe1997}
    M. Ackermann \emph{et al.}, Astropart. Phys. \textbf{22}, 339 (2005).

\bibitem{rice}
    I. Kravchenko \emph{et al.}, Phys. Rev. D \textbf{73}, 082002
(2006). The value of the upper limit is taken from Figure 19. S. Hussain
(private communication).  

\bibitem{icecubesensitivity}
    J. Ahrens \emph{et al.}, Astropart. Phys. \textbf{20}, 507 (2004).

\bibitem{target}
    R. Engel \emph{et al.}, in \textit{Proceedings of the 27th
International Cosmic Ray Conference}, Hamburg, Germany, 1381 (2001).

\bibitem{ams}
    J. Alcaraz \textit{et al.} (AMS Collaboration), Phys. Lett. B \textbf{490}, 27 (2000).

\bibitem{bess}
    T. Sanuki \textit{et al.}, Astrophys. J. \textbf{545}, 1135 (2000).

\bibitem{bess-tev}
    S. Haino \textit{et al.}, Phys. Lett. B \textbf{594}, 35 (2004).

\bibitem{dpmjet}
    S. Roesler, R. Engel, and J. Ranft, in \textit{Proceedings of the 27th
International Cosmic Ray Conference}, Hamburg, Germany, \textbf{1}, 439 (2001);
Phys. Rev. D \textbf{57}, 2889 (1998).

\end{thebibliography}
\end{document}